\documentclass[useAMS,usenatbib]{mn2e} \usepackage{graphicx}

\title[Methanol maser catalogue: 6$^{\circ}$ to 20$^{\circ}$]{The 6-GHz methanol multibeam maser catalogue II: Galactic longitudes 6$^{\circ}$ to 20$^{\circ}$}

\author[Green et al.]
       {J. A. Green,$^1$\thanks{E-mail:james.green@csiro.au} J. L.
Caswell,$^1$ G. A. Fuller,$^2$ A. Avison,$^{2}$ S. L. Breen,$^{3,1}$  \newauthor
S. P. Ellingsen,$^3$ M. D. Gray,$^{2}$ M. Pestalozzi,$^{5}$ L. Quinn,$^2$ M. A.
Thompson,$^4$ \newauthor M. A. Voronkov$^1$ \\ 
$^{1}$ CSIRO Astronomy and Space Science, Australia Telescope
National Facility, PO Box 76, Epping, NSW 2121, Australia; \\ $^{2}$
Jodrell Bank Centre for Astrophysics, Alan Turing Building, School of Physics and Astronomy, University of
Manchester,\\ Manchester, M13 9PL, UK; \\ 
$^{3}$ School of Mathematics and Physics, University of Tasmania, Private
Bag 37, Hobart, TAS 7001, Australia; \\ $^{4}$
Centre for Astrophysics Research, Science and Technology Research
Institute, University of Hertfordshire, College Lane, \\Hatfield, AL10
9AB, UK; \\ $^{5}$ INAF/IFSI, via del Fosso del Cabaliere 100, I-00133, Roma \\}

\date{Accepted 2010 July 15.  Received 2010 July 13; in original form 2010 May 6}

\pagerange{\pageref{firstpage}--\pageref{lastpage}} \pubyear{XXXX}
\begin{document} \maketitle

\label{firstpage}

\begin{abstract}
We present the second portion of an unbiased survey of the Galactic plane for 6668-MHz methanol masers. 
This section of the survey spans the longitude range 6$^{\circ}$  to 20$^{\circ}$.  We report the detection of 119 maser
sources, of which 42 are new discoveries. The masers are tightly constrained to the Galactic plane, with only four outside a latitude range of $\pm$1$^{\circ}$.
This longitude region includes the brightest known 6668-MHz methanol
maser, 9.621+0.196, as well as the two brightest newly discovered
sources in the southern survey as a whole. We list all the sources associated with the 3-kpc arms within $\pm$15$^{\circ}$ longitude and consider further candidates beyond 15$^{\circ}$ longitude. We identify three new sources associated with the Galactic bar and comment on the density of masers in relation to the bar orientation.

\end{abstract}


\section{Introduction}
The Methanol Multibeam (MMB) survey is a project to survey the entire Galactic plane for 6668-MHz methanol masers  \citep{green09a}. This species of maser is one of the brightest observed, it is widespread throughout the Galaxy, and it exclusively traces high-mass stars in an early stage of their formation \citep{pestalozzi02b, minier03, xu08}. This makes the 6668-MHz methanol maser a powerful tool for understanding the processes of high-mass star formation and studying the spiral arm structure of our Galaxy \citep[e.g.][]{reid09,rygl10}. By surveying in an unbiased manner, with uniform sensitivity, the MMB will establish a definitive catalogue for future studies. More than 60\% of the Galactic plane has now been observed with the Parkes Radio Telescope and detections followed up to yield accurate positions ($\le$0.4 arcsec).
The catalogue is being released sequentially, as data reduction and follow-up observations are completed. The first region to be released covered Galactic longitudes 345$^{\circ}$ to 6$^{\circ}$ \citep{caswell10mmb1}, with a focus on recognising masers associated with the Galactic centre. The current paper extends the Galactic centre region to a longitude of 20$^{\circ}$,  and represents the remaining sources north of the Galactic centre which have high resolution positions from the Australia Telescope Compact Array (ATCA). For sources further north we are obtaining positions with the Multi-Element Radio Linked Interferometer Network (MERLIN) and the Very Large Array (VLA) and these will be presented in a subsequent paper. 

A number of targeted searches have previously been made in the 6$^{\circ}$ to 20$^{\circ}$ longitude region presented here, with the targets consisting of masers of other methanol transitions and other molecular species (e.g. 12-GHz methanol and 1665/1667-MHz hydroxyl masers), as well as various star formation indicators (e.g. IRAS colour selected ultra-compact H{\sc ii} regions). The targeted observations include the first search for methanol at 6668-MHz \citep{menten91} that led to the discovery of maser emission in this transition. The other major targeted observations preceding the MMB survey were: \citet{schutte93, walt95, caswell95a, walsh97, walsh98, slysh99, szymczak00, beuther02}. The combination of these observations resulted in 74 
sources known in this region prior to the MMB survey \citep{pestalozzi05}. The positions for 45 of these were presented in \citet{caswell09a}. During the course of the survey three more sources have been detected (\citealt{ellingsen07} and \citealt{cyganowski09}).

Of special interest, this survey region includes the remaining sources regarded by \citet{green09b} as a likely population associated with the enigmatic 3-kpc arms \citep{vanwoerden57,oort58,bania77,bania80,lockman80}. The presence of 6668-MHz methanol masers in this interior Galactic feature conclusively demonstrated the presence of high-mass star formation, in both the well known near-side portion, and the recently discovered far-side portion \citep{dame08}. Understanding the nature and structure of this part of the Milky Way is key to understanding the structure and dynamics of the inner Galaxy.

\section{Methanol multibeam survey parameters and equipment}
A full review of the survey techniques is given in \citet{green09a}, so only a brief description is presented here together with details specific to the 6$^{\circ}$ to 20$^{\circ}$ longitude region. The survey has been conducted with a purpose-built 7-beam receiver on the Parkes radio telescope together with subsequent ATCA observations (in the 6-km configurations) to provide high resolution positions (positional accuracy of $\sim$0.4 arcsec). At Parkes, blocks of the Galactic plane 2$^\circ$ in longitude by 4$^\circ$ in latitude were initially raster scanned. Maser sources in the resulting data cubes were first identified using an automated routine and then visually inspected. Once the high resolution positions were determined with the ATCA, seven-minute pointed Parkes observations, referred to as MX observations, were subsequently obtained towards the maser sources (including two sources known to exist outside the MMB latitude range). The Parkes beamwidth at 6668-MHz is 3.2 arcmin and the typical 1 $\sigma$ noise level in the survey cubes was 0.17 Jy. The MX
observations typically reached noise levels of 0.07 Jy. The spectral resolution at Parkes was 0.11\,km\,s$^{-1}$. The latitude range of the MMB was $\pm$2$^{\circ}$, which was adequate for this region of the Galactic plane, because only four sources are known to exist outside $\pm$1$^{\circ}$ and only two of these lie more than 2$^{\circ}$ from the plane. 

The survey cubes for 6$^{\circ}$ to 14$^{\circ}$ longitude were observed over the period 2006 April to 2007 August, whereas 14$^{\circ}$ to 20$^{\circ}$ longitude were observed solely in 2007 August. Twelve of the 32 scans in the 8$^{\circ}$ to 10$^{\circ}$ longitude region were repeated in 2007 November due to higher than average noise in the originals. The MX observations for the 6$^{\circ}$ to 20$^{\circ}$ longitude region were taken in 2008 March, 2008 August and 2009 March.

As detailed in \citet{green09a}, the velocity coverage was designed to cover all the known CO emission of \citet{dame01}, with multiple velocity settings where needed (single setting coverage was $\sim$180\,km\,s$^{-1}$). For the longitude range covered in this paper, two central velocity settings were required: between 6$^{\circ}$ and 10$^{\circ}$ these were centred at $+$135 km\,s$^{-1}$ and $-$10 km\,s$^{-1}$; and between 10$^{\circ}$ and 20$^{\circ}$ they were centred at $+$145 km\,s$^{-1}$ and 0 km\,s$^{-1}$.

\section{Survey results}
Results are presented in Table \ref{resotable}: column one is the source name (Galactic longitude
and latitude); columns two and three are Right Ascension and Declination in J2000 co-ordinates from ATCA observations; columns four and five are the minimum and maximum velocities of emission (the most extreme values observed in any observations); columns six and seven are the MX peak velocities and flux densities; columns eight and nine are the peak velocities and flux densities observed in the survey cubes; and the final column provides extra information on the date of the ATCA observation (if it was observed by us) or a reference for comparable observations made previously. All velocities are with respect to the radio convention of the local standard of rest (LSR). For sources in close proximity, individual source velocity ranges have been determined from inspection of the higher spatial resolution ATCA data. Parkes spectra for the sources are given in Figure \ref{spectra}. These are all MX spectra with the exception of 14.230-0.509, which is the survey cube spectrum (the source is variable and was not detectable in the MX observations as discussed in Section\,\ref{srcnotes}). Typical physical sizes ($\le$0.03\,pc) and distances ($\ge$3\,kpc) are such that maser spots associated with an individual exciting star (or binary) are spread over a diameter of less than 2 arcsec \citep{caswell97}; as such we treat emission features separated by $\ge$2 arcsec as different maser sites \citep{caswell09a, caswell10mmb1}. All our sources are separated by at least 10 arcsec, with the exception of the two pairs 18.733-0.224 with 18.735-0.227, and 19.472+0.170 with 19.472+0.170n. For completeness, we list the two additional sources known to exist outside the latitude range of the MMB survey \citep{pestalozzi05} and give measurements from our targeted MX observations.

For the 6$^{\circ}$  to 20$^{\circ}$  longitude region we detected 119 sources, corresponding to an average of 8 sources per degree of longitude, with 42 new detections. 

\subsection{Remarks on sites of maser emission.}\label{srcnotes}
Here we provide additional details on notable sources and sources with confused spectral structure. This includes: 11 sources which are associated with the 3-kpc arms \citep{green09b}; 4 sources with relatively large latitudes; 6 with known variability; and 3 with velocity ranges wider than 16\,km\,s$^{-1}$. 
The majority of sources within longitudes 6$^{\circ}$ to 20$^{\circ}$ are located within the solar circle (inferred from their positive LSR velocities) and therefore have a near-far ambiguity for kinematic distance estimates. Where possible we remark on sources which have been allocated a near or far distance in the literature.

\subparagraph{6.189-0.358} This is the brightest (229 Jy) new detection of the MMB survey and is associated with the near 3-kpc arm.

\subparagraph{6.368-0.052, 7.601-0.139 and 7.632-0.109} These three new sources all have high velocities (144.1\,km\,s$^{-1}$, 154.7\,km\,s$^{-1}$ and 157.0\,km\,s$^{-1}$ respectively) and are potentially associated with the Galactic bar. For the closely spaced pair, 7.601-0.139 incorporates the features between 155 and 157 \,km\,s$^{-1}$, whilst 7.632-0.109 incorporates the features outside these velocities. All three sources may be linked to the known high velocity H{\sc i} at these longitudes, which extends from the plane to high latitudes \citep[see for example][]{mcclure05}. 6.368-0.052 displays a feature of 0.2 Jy at 136\,km\,s$^{-1}$ in both the MX taken 2008 March and the MX taken 2009 March, however it has not been confirmed with the ATCA.

\subparagraph{6.539-0.108, 6.588-0.192 and 6.610-0.082} We have detected a new source amongst two previously known sources. The new source, 6.588-0.192, has three blended peaks between 4 and 6\,km\,s$^{-1}$. 6.539-0.108 has a clear feature at 13.5\,km\,s$^{-1}$ and 6.610-0.082 has a clear feature at 0.8\,km\,s$^{-1}$, both known from previous observations. Additionally we detect weak emission between 6 and 7\,km\,s$^{-1}$ in the spectrum of 6.539-0.108 and weak emission ($<$0.9\,Jy) at 5-6\,km\,s$^{-1}$ and $\sim$10\,km\,s$^{-1}$ in the spectrum of 6.610-0.082. The features in both spectra near 6\,km\,s$^{-1}$ may partially be a sidelobe response of the new source 6.588-0.192. A fourth source may exist in the region, but it could not be resolved with the ATCA, and would require future VLBI observations. LSR velocities near zero at this Galactic longitude imply heliocentric distances of either a few kpc or beyond 10 kpc. \citet{downes80b} ascribe the associated H{\sc ii} region 6.553-0.095 to the far kinematic distance based on formaldehyde absorption. They also speculated that it might be in the 3-kpc arm, but this seems unlikely since the velocities do not correspond with either of the arms as defined in \citet{dame08}.

\subparagraph{6.795-0.257} This source has a wide velocity range of 19.3\,km\,s$^{-1}$ with the features between 25\,km\,s$^{-1}$ and 30\,km\,s$^{-1}$ probably separated from the bright feature at $\sim$14\,km\,s$^{-1}$ by approximately one arcsecond. Overall the features increased in flux density between the survey cube observation and the MX (2009 March).

\subparagraph{7.166+0.131} This is a new source to the survey and clearly associated with the far 3-kpc arm. It also exhibits a wide velocity range of 16.5\,km\,s$^{-1}$.

\subparagraph{8.139+0.226} This source, when first discovered by \citet{schutte93}, had a peak flux density of $\sim$15 Jy. A position was obtained in 2000 by \citet{caswell09a} with a peak flux density of only 3.5 Jy. The survey cube spectrum (taken 2006/2007) showed a higher peak flux density of 5.2 Jy. The MX taken in 2009 showed a peak flux density of over twice this at 11.4 Jy. On the basis of a lack of formaldehyde absorption, \citet{wink82} suggest the associated compact H{\sc ii} region is most likely at the near kinematic distance. 
 
\subparagraph{8.669-0.356 and 8.683-0.368} The spectra of this close pair of sources contain a feature at $\sim$36\,km\,s$^{-1}$ which is likely to be variable and could not be clearly attributed to either one of the sources. 8.669-0.356 comprises only the small feature at $\sim$39\,km\,s$^{-1}$, whilst 8.683-0.368 contains all other features seen in the spectrum. \citet{downes80b} ascribe the associated H{\sc ii} region to the near kinematic distance based on a lack of formaldehyde absorption. 

\subparagraph{8.832-0.028} This is a bright new source (159 Jy) also recently positioned by \citet{xu09b} with the ATCA. It is associated with the near 3-kpc arm.

\subparagraph{8.872-0.493} This new source was also recently positioned by \citet{xu09b} with the ATCA. It is potentially associated with the molecular cloud 8.9-0.5 (velocity of 12\,km\,s$^{-1}$) from  \citet{solomon87}, which was claimed by the authors to have kinematics matching those of the near 3-kpc arm. However, the kinematics do not match an extrapolation of the arm based on 
 \citet{cohen76} nor the more recent definition of \citet{dame08}. Applying current kinematic models, the velocity of 8.872-0.028 ($\sim$23\,km\,s$^{-1}$) would place it in either the Carina-Sagittarius or Perseus arms. If the maser is associated with the 8.9-0.5 molecular cloud (with a 10\,km\,s$^{-1}$ unusual velocity), then that would place it more likely in the Perseus, rather than the Carina-Sagittarius arm. In either case the source is unlikely to be associated with the near 3-kpc arm.

\subparagraph{9.621+0.196 and 9.619+0.193} This famous source and its companion have been extensively studied since discovery in 1991 \citep{menten91, norris93, walsh97, walsh98, phillips98,  caswell09a}. There has also been extensive monitoring of 9.621+0.196 by \citet{caswell95d, goedhart03,goedhart04,walt09, vlemmings09}, with \citet{goedhart03} detecting periodic flares (see Section\,\ref{varisection} for details). The weaker source, 9.619+0.193, probably has just two features. Several authors have postulated the pair lie within the 3-kpc arm \citep[e.g.][]{caswell95a, hofner94, green09b} and recent trigonometric parallax measurements of 12 GHz methanol \citep{sanna09} established a heliocentric distance of 5.2$\pm$0.6 kpc (corresponding to a Galactocentric distance consistent with 3-kpc arm estimates).

\subparagraph{9.986-0.028} This source consists of a group of three strong ($>$10\,Jy) features between 40\,km\,s$^{-1}$ and 45\,km\,s$^{-1}$ (two of which are blended in the Parkes spectrum), an outlying strong feature at 47\,km\,s$^{-1}$ and other weaker ($<$10\,Jy) emission between 40\,km\,s$^{-1}$ and 52\,km\,s$^{-1}$. The 47\,km\,s$^{-1}$ feature was the brightest at 35\,Jy in the original \citet{schutte93} Hartebeesthoek observations, but had decreased to 28\,Jy in the ATCA observations made in 2000 by \citet{caswell09a}. In our observations this feature is stable between the survey cube (2007 January) and MX (2009 March) observations with a peak flux density of $\sim$27\,Jy. Meanwhile the feature at 42\,km\,s$^{-1}$ which had shown a steady increase from $\sim$15\,Jy to 26\,Jy between the  \citet{schutte93} and \citet{caswell09a} observations now dominates the spectrum. Our survey cube and MX observations find a peak flux density of $\sim$70\,Jy. The 43.5\,km\,s$^{-1}$ feature decreased from 32\,Jy in 1993 to 20\,Jy in 2000, increased to 24 Jy in the survey cube and decreased to 20\,Jy in the MX. The 40.5\,km\,s$^{-1}$ feature has been stable at $\sim$25\,Jy across all the observations with the exception of the survey cube where it flared to 38\,Jy.

\subparagraph{10.205-0.345, 10.287-0.125, 10.299-0.146, 10.323-0.160 and 10.342-0.142 } These are the first of four groups of sources which are in the direction of the W31 complex. The close proximity of all these sources leads to confusion in the Parkes spectra, however the high spatial resolution of the ATCA observations allows identification of the different source features: 10.205-0.345 has emission between 5 and 11\,km\,s$^{-1}$ with a peak at 7.2\,km\,s$^{-1}$; 10.287-0.125 contains the three features with velocities less than 5\,km\,s$^{-1}$; 10.299-0.146 is the small feature at $\sim$20\,km\,s$^{-1}$;  10.323-0.160 is the feature at 6\,km\,s$^{-1}$ and the double feature between 9 and 13\,km\,s$^{-1}$; 10.342-0.142 exists between 7 and 17\,km\,s$^{-1}$, but not including the bright peaks of the previous mentioned source. 10.323-0.160 has been monitored for variability by \citet{goedhart04}. 10.287-0.125 and 10.342-0.142 are associated with Extended Green Objects \citep{cyganowski09}.  \citet{wilson84} detected methanol absorption at 23 GHz. 
  
\subparagraph{10.320-0.160 and 10.356-0.148} These sources, also in the direction of W31, have emission at higher velocities:
10.320-0.160 comprises the features between 35 and 40\,km\,s$^{-1}$; 10.356-0.148 the features between between 49 and 54\,km\,s$^{-1}$.

\subparagraph{10.444-0.018, 10.472+0.027 and 10.480+0.033} Although close to W31 spatially, the high velocities of this trio of sources separates them from the complex. 10.472+0.027 and 10.480+0.033, separated by $\sim$6 arcmin, and originally identified as 10.47+0.03, were highlighted by \citet{caswell95d} as extremely variable and have been monitored for variability by \citet{goedhart04}. The sources have a very wide range of velocity emission which is comparable to the nearby source 10.444-0.018. 

\subparagraph{10.627-0.384 and 10.629-0.333} This is the fourth group of sources which have been loosely associated with W31. 10.627-0.384 includes the small feature on the edge of the absorption at approximately -6\,km\,s$^{-1}$, but not the outlying features. Absorption was also evident in the Parkes spectrum shown by \citet{caswell95a}. 10.629-0.333 is kinematically associated with the near 3-kpc arm as defined by \citet{dame08}.  Although marginally offset in peak velocity, the close spatial association of 10.627-0.384 together with its emission at negative velocities, implies this source is also associated. The other 3 groups of sources previously mentioned are offset in position and velocity from 10.627-0.384 and 10.629-0.333, so do not appear to be associated with the near 3-kpc arm as defined by \citet{dame08}. This implies that 10.629-0.333 and 10.627-0.384 may be a chance spatial alignment with W31 and not actually associated. Conversely if the whole W31 complex is associated with the near 3-kpc arm, as was suggested by \citet{downes80b}, the 3-kpc arm kinematics differ from that observed by \citet{dame08}. \citet{fish03} argues that 10.629-0.333 is unlikely to be at the far distance based on HI absorption observations towards an associated compact H{\sc ii} region. 

\subparagraph{10.724-0.334} This source is associated with the near 3-kpc arm. It has significantly brightened from 1.5 Jy in the survey cube observation to 5 Jy in the MX.

\subparagraph{10.822-0.103} This source had three features in the survey cube observation, between 68\,km\,s$^{-1}$ and 74\,km\,s$^{-1}$. The ATCA data detected all three, but at minimal signal-to-noise. The MX taken 2009 March only detected one feature at a velocity of 69\,km\,s$^{-1}$.

\subparagraph{10.886+0.123} Both the survey cube and MX data (2008 March) show consistency in the shape of the spectrum and the peak flux densities of the features, although the ATCA spectrum had flux densities a factor of two lower.

\subparagraph{11.034+0.062} The 25\,km\,s$^{-1}$ feature seen in the spectrum is a sidelobe response to 10.958+0.022.

\subparagraph{11.109-0.114} This new source to the survey was also detected by \citet{ellingsen07}, but listed as 11.15-0.14. 

\subparagraph{11.903-0.102 and 11.904-0.141} Of this pair of sources 11.904-0.141 accounts for the features between $\sim$40 and 45\,km\,s$^{-1}$ and 11.903-0.102 for the lower velocity features. No variability was present between the observations of \citet{caswell95a} and \citet{caswell97}, however our observations find that for 11.903-0.102 the original peak feature at 36\,km\,s$^{-1}$ has increased from 1.8\,Jy \citep{caswell97} to 4.5\,Jy. Additionally the feature at 33.8\,km\,s$^{-1}$ has increased from 1.5\,Jy \citep{caswell97} to $\sim$11.5\,Jy.  

\subparagraph{11.936-0.150} This source only contains the features seen between 45 and 50\,km\,s$^{-1}$. Features $<$45\,km\,s$^{-1}$  are side-lobe responses to 11.904-0.141.

\subparagraph{11.936-0.616} This source is associated with an Extended Green Object  \citep{cyganowski09}. \citet{solomon87} attribute the associated molecular cloud, 12.00-0.60, to the near kinematic distance based on a velocity-linewidth argument. 

\subparagraph{12.025-0.031} This is a known source clearly associated with the far 3-kpc arm.

\subparagraph{12.181-0.123, 12.199-0.033, 12.202-0.120, 12.203-0.107 and 12.209-0.102} This complex cluster of 5 sources is divided as: 12.181-0.123 which consists of the two features at $\sim$30\,km\,s$^{-1}$; 12.202-0.120 which comprises the features between 20 and $\sim$32.5\,km\,s$^{-1}$; 12.203-0.107 which is the feature peaking at 20.5\,km\,s$^{-1}$; 12.209-0.102 which is composed of the weak features between 15 and $\sim$23\,km\,s$^{-1}$; 12.199-0.033, although spatially close, is well separated in velocity. 

\subparagraph{12.625-0.017} The brightest feature of this maser at 21.6\,km\,s$^{-1}$ has increased from $\sim$10 Jy (2007 August) to almost 25 Jy (2008 March).

\subparagraph{12.681-0.182} This source has previously been seen to vary from maximum flux density to minimum flux density by a factor of 1.5 over a three month period \citep{caswell95d} and was a source monitored for variability by \citet{goedhart04}, who found that all maser peaks exhibited a simultaneous variation with a quasi-periodic time-scale of 307$\pm$60 days. We made three observations, the original survey cube in 2007 August and two MX observations taken 2008 March and 2009 March. The 2007 August and 2008 March spectra show consistent spectral features (with flux densities comparable to within 5\%) with the exception of the two features at 59 and 60\,km\,s$^{-1}$, which decreased by $\sim$20\%. In the 2009 March spectrum the features at 57.5, 59 and 60\,km\,s$^{-1}$ had all decreased from the 2007 August flux densities by 12\%, 60\% and 46\% respectively.  

\subparagraph{12.889+0.489}  The features of this source were seen to vary from maximum flux density to minimum flux density by a factor of 1.3 over an eight month period \citep{caswell95d} and it was then monitored for variability by \citet{goedhart04} and \citet{goedhart09}. It has been shown to have a short period of only 29.5 days. The MMB survey cube observations have a peak flux density of 78.9 Jy (2007 August) and the MX 68.9 Jy (2008 March). See also Section\,\ref{varisection}.

\subparagraph{12.909-0.260} This source is associated with W33 and has been monitored for variability by \citet{goedhart04}. The authors found the 39.4\,km\,s$^{-1}$ and 39.8\,km\,s$^{-1}$ features showed little variation, whilst the 35.9\,km\,s$^{-1}$ feature had a monotonic increase from 12 to 20 Jy over their duration of their observations. We find a peak flux density of 245 Jy at 39.9\,km\,s$^{-1}$ in the survey cube rising to 269 Jy in the later MX (2008 March). Additional MX observations in 2009 March found 210 Jy (at 39.4\,km\,s$^{-1}$) and 250 Jy (at 39.9\,km\,s$^{-1}$). The feature at 35.9\,km\,s$^{-1}$ had a flux density of 25 Jy at the time of the survey cube, 22.5 Jy in 2008 March MX, and 22 Jy in 2009 March MX. 
The nearby object listed as 12.79-0.19 in \citet{menten91} and not detected by either  \citet{caswell95a} or \citet{caswell09a} was again not detected in a targeted MX observation. Clear absorption of $\sim$2 Jy was found in the Parkes spectrum, but no detectable spectral structure indicative of maser emission. In agreement with \citet{voronkov10}, this location is about 7 arcmin from 12.909-0.260 and the 5 Jy peak flux density listed by  \citet{menten91} implies 12.79-0.19 was a sidelobe response to 12.909-0.260.

\subparagraph{13.696-0.156}  This is a new source detected by the MMB survey and is clearly associated with the far 3-kpc arm.

\subparagraph{14.101+0.087} This source is associated with the near 3-kpc arm and a position was recently obtained by \citet{xu09b}.

\subparagraph{14.230-0.509} This source was observed as a single feature peak of 3.6 Jy at 25.3\,km\,s$^{-1}$ in the survey cube observations. 
It was then undetectable in both the 2008 March MX and the ATCA observation taken in 2008 October. Fortunately it was detected again with the ATCA in 2009 January and a position was successfully obtained, with a peak flux density of 0.4 Jy at the original velocity of 25.3\,km\,s$^{-1}$. It was then undetected in a subsequent MX taken 2009 March.

\subparagraph{14.521+0.155} This is a new source associated with the near 3-kpc arm. A feature of 0.5\,Jy  at -2\,km\,s$^{-1}$ was seen in the MX observation (2009 March) which was not present in the prior survey cube observation.

\subparagraph{14.991-0.121} This source includes two weak features at 52.5\,km\,s$^{-1}$ and 54\,km\,s$^{-1}$.

\subparagraph{15.034-0.677} This source is associated with the H{\sc ii} region in M17 \citep{caswell97}. The brighter of the two features in this source was observed in the survey cubes to be stronger than any previous observations with a peak flux density of 51.6 Jy. The weaker of the two features has remained constant at $\sim$10 Jy. 

\subparagraph{15.607-0.255} This source was originally observed at 0.85 Jy, but faded to 0.4 Jy in an MX taken 2008 March and then was not detectable ($<$0.2 Jy) in the MX taken 2008 August. It was however detected again in 2008 October with the ATCA (with a peak flux density of 0.4\,Jy).

\subparagraph{16.302-0.196} The velocity range of emission is nominally 46.9\,km\,s$^{-1}$ to 53.6\,km\,s$^{-1}$, but there is a weak spectral feature at $\sim$41\,km\,s$^{-1}$ in one MX, which was not seen in the original survey cube observation, or the ATCA observation.

\subparagraph{16.585-0.051} The 59\,km\,s$^{-1}$ feature of this source was seen to be steadily decreasing in flux density over a period of 6 months by \citet{caswell95a}. We find this feature is marginally weaker at 18\,Jy (compared to $\sim$20\,Jy in Caswell et al.). The feature at 64\,km\,s$^{-1}$ has remained stable at $\sim$15\,Jy, but the 62\,km\,s$^{-1}$ feature has significantly increased from 18\,Jy to 37\,Jy.  In the current observations we do not detect ($<$0.15 Jy) the two weak features at 52.8\,km\,s$^{-1}$ and 54.2\,km\,s$^{-1}$ seen by Caswell et al., instead finding a range of emission from 56.5 to 69.5\,km\,s$^{-1}$. However for completeness we list the full velocity range in Table \ref{resotable}.

\subparagraph{16.864-2.159 and 17.021-2.403} Both these sources lie outside the latitude range of the MMB survey, but were re-observed with MX observations for completeness. 16.864-2.159 was originally discovered by \citet{caswell95a} in a targeted search towards OH masers and has since been observed by \citet{walsh97, walsh98} and \citet{szymczak00}. Its spectrum has not changed over the $>$10 year period of these observations and our own. 17.021-2.403 was discovered by \citet{szymczak00} in a targeted search of IRAS sources and our observations expand the velocity range of the source with the detection of a weak ($<$1\,Jy) feature at 18\,km\,s$^{-1}$. The feature at 21\,km\,s$^{-1}$ has also increased significantly from the original observation of $\sim$1\,Jy to 4\,Jy. 

\subparagraph{16.976-0.005} This is a new weak source with a peak flux density of 0.7\,Jy at 6.6\,km\,s$^{-1}$. 

\subparagraph{17.638+0.157} This source was discovered in 1991 by \citet{macleod92b} with a peak flux density of 25\,Jy at 21\,km\,s$^{-1}$. It was observed again in 1992 by \citet{caswell95a} with a flux density of 25\,Jy and in 1999 by \citet{szymczak00} with a reduced flux density of 20\,Jy.  A position was determined with the ATCA in 2003 \citep{caswell09a}, finding a significantly further reduced flux density of 9.5\,Jy. Our survey cube observations in 2007 August found an increase again with a peak flux density of 35\,Jy. The MX in 2009 March recorded a peak flux density comparable to the original value of 25\,Jy. 

\subparagraph{18.073+0.077} This source contains two bright features separated by approximately 6\,km\,s$^{-1}$. 

\subparagraph{18.159+0.094} The relative intensity of the two strong features at 58 to 59\,km\,s$^{-1}$ has varied, with the 58.4\,km\,s$^{-1}$ feature remaining relatively constant at $\sim$8.5 Jy, but the feature at 59\,km\,s$^{-1}$ decreasing from 10.3 Jy in the survey cube (2007 August) to just shy of 2 Jy in the MX (2009 March). In addition to the major features, the MX detected a small feature ($\sim$1 Jy) at 54.5\,km\,s$^{-1}$, which was only marginally detected at 0.7 Jy in the survey cube and not detected in the ATCA observations.

\subparagraph{18.440+0.045} This source has noticeable changes in its structure with the peak moving from $\sim$58\,km\,s$^{-1}$ (2007 August) to $\sim$62\,km\,s$^{-1}$ (2008 August) and a new feature appearing at $\sim$60\,km\,s$^{-1}$. The feature at 49\,km\,s$^{-1}$ is a sidelobe of  18.460-0.004.

\subparagraph{18.661+0.034 and 18.667+0.025} These are a closely spaced pair of sources, previously listed as one known site, but now distinguished as two sites. 18.661+0.034 extends to the higher velocities. 18.667+0.025 is mainly the features at 77\,km\,s$^{-1}$ and 80.5\,km\,s$^{-1}$ and is associated with an Extended Green Object \citep{cyganowski09}.

\subparagraph{18.735-0.227 and 18.733-0.224} These sources are a closely spaced pair of new detections. The bright peak feature at 38\,km\,s$^{-1}$ belongs to 18.735-0.227, but the 3 other peaks are 18.733-0.224.

\subparagraph{18.888-0.475} This new source was also recently detected by \citet{cyganowski09} associated with an Extended Green Object.

\subparagraph{19.009-0.029} This new source was also recently detected by \citet{ellingsen07}, listed as 18.99-0.04. It was seen offset from a GLIMPSE source by 1.5 arcmin and is unlikely to be physically associated. 

\subparagraph{19.249+0.267} This source does not include the feature seen at  25\,km\,s$^{-1}$ which is a sidelobe of 19.365-0.030.

\subparagraph{19.472+0.170n, 19.472+0.170 and 19.486+0.151} Two sources separated by less than 4 arcsec, are distinguished by an `n' identifying the source with the more northerly declination (previously the southern site was instead identified with an `sw' e.g. \citealt{caswell09a}, but we adopt an `n' in accord with the Galactic centre region results of \citealt{caswell10mmb1}). The northern site contains the features at 18.5, 22 and 23\,km\,s$^{-1}$ the southern site, the features between 13 and 16\,km\,s$^{-1}$. 19.486+0.151 comprises the other features at 20.5\,km\,s$^{-1}$, 24\,km\,s$^{-1}$ and beyond to higher velocities.  

\subparagraph{19.609-0.234} This source consists of two main features, one at 36\,km\,s$^{-1}$ and one at 40\,km\,s$^{-1}$. The 36\,km\,s$^{-1}$ feature had been the strongest feature, at 0.4 Jy, in the 1992 discovery spectrum \citep{caswell95d}. However this feature was a marginal detection in the survey cube data (2007 August).  Fortunately the weak feature of 0.25 Jy at 40\,km\,s$^{-1}$ in 1992 flared to 1 Jy in the survey cube, allowing an ATCA position measurement in 2007 July (0.4 Jy peak). The 36\,km\,s$^{-1}$ feature rose to 0.5 Jy in an MX in March 2008, but was undetectable in an MX in March 2009. The 40\,km\,s$^{-1}$ feature had a peak flux density of 0.65 Jy in the MX measurement of 2009 March (and is shown in Figure \ref{spectra}). \citet{kolpak03, anderson09a, roman09} ascribe this source to a far kinematic distance based on an absence of HI self-absorption, whilst \citet{downes80b} ascribe it to a near kinematic distance based on an absence of formaldehyde absorption.

\subparagraph{19.612-0.120 and 19.612-0.134} As identified by \citet{walsh98}, in addition to the main source 19.612-0.120 which has several features between 49 and 61\,km\,s$^{-1}$,  there is a narrow offset source peaking at 52-53\,km\,s$^{-1}$. Both \citet{kolpak03} and \citet{anderson09a} ascribe the associated H{\sc ii} region to a far kinematic distance based on HI absorption and HI self-absorption, in contrast to \citet{downes80b}, who ascribe it to a near kinematic distance based on the absence of formaldehyde absorption. 

\subparagraph{19.667+0.114} This source peaked in emission at 16.3\,km\,s$^{-1}$ in the survey cube observation, but peaked at 14.2\,km\,s$^{-1}$ in the MX observation, with the 14\,km\,s$^{-1}$ feature doubling in strength from the MX observation from 1.1 Jy to 2.2 Jy. The features either side of the 16.3\,km\,s$^{-1}$ feature have also increased considerably in flux density (whilst the original 16.3\,km\,s$^{-1}$ feature has remained approximately the same).

\subparagraph{19.884-0.534} This source has been identified as a far-side object from the presence of formaldehyde absorption (at a level of 7$\sigma$) by \citet{sewilo04}. In contrast \citet{roman09} identify it as a near-side object based on the `on-off' HI self absorption technique, as do \citet{solomon87} based on a line-width to distance relation.

\begin{table*} \centering \caption{\small Positions and parameters of methanol masers. The references are: W98: \citet{walsh98}, B02: \citet{beuther02}, X09: \citet{xu09b}, C2009: \citet{caswell09a}. Position is from non-bracketed reference.} 
\begin{tabular}{lcrcccrcrl}
\hline

\multicolumn{1}{c}{Source Name} & \multicolumn{2}{c}{Equatorial
Coordinates} & \multicolumn{2}{c}{Velocity range} & \multicolumn{2}{c}{MX
data} & \multicolumn{2}{c}{Survey Cube data} & \multicolumn{1}{l}{Position}\\

\ (~~~l,~~~~~~~b~~~)    &       RA(2000)        &       Dec(2000)       &
$\rm V_{L}$&$\rm V_{H}$ &  $\rm V_{pk}$(MX)     &  $\rm S_{pk}$(MX)  &
$\rm V_{pk}$(SC) & $\rm S_{pk}$(SC)     &   Refs,
epoch    \\
\ (~~~$^\circ$~~~~~~~$^\circ$~~~) & (h~~m~~~s) & (~$^\circ$~~ '~~~~") &
\multicolumn{2}{c}{(km\,s$^{-1}$ )} & (km\,s$^{-1}$ ) &  (Jy) & (km\,s$^{-1}$ ) & (Jy) & \\

\hline
06.189$-$0.358  &  18 01 02.16  &  -23 47 10.8  & -37.5 & -27.1 & -30.2 & 228.57 & -30.2 & 221.60 &  2007JUL19\\
06.368$-$0.052  &  18 00 15.82  &  -23 28 43.8  & 141.0 & 147.8 & 144.1 & 1.50 & 144.1 & 1.49 &  2007JUL22 \\
06.539$-$0.108  &  18 00 50.86  &  -23 21 29.8  & 12.7 & 13.8 & 13.1 & 0.60 & 13.4 & 0.50 &  C2009 \\
06.588$-$0.192  &  18 01 16.09  &  -23 21 27.3  & 3.5 & 7.0 & 5.1 & 8.01 & 5.0 & 7.70 &  2007JUL21 \\
06.610$-$0.082  &  18 00 54.03  &  -23 17 03.1  & -6.6 & 7.5 & 0.8 & 23.40 & 0.8 & 21.20 &  C2009 \\
06.795$-$0.257  &  18 01 57.75  &  -23 12 34.9  & 12.1 & 31.4 & 16.3 & 91.07 & 16.3 & 55.08 &  C2009 \\
06.881+0.093  &  18 00 49.38  &  -22 57 42.6  & -3.8 & -1.5 & -2.3 & 3.12 & -2.1 & 3.25 &  2007JUL21 \\
07.166+0.131  &  18 01 17.48  &  -22 41 44.0  & 74.5 & 91.0 & 85.7 & 2.58 & 85.7 & 2.47 &  2007JUL22\\
07.601$-$0.139  &  18 03 14.43  &  -22 27 00.9  & 151.0 & 156.5 & 154.6 & 8.69 & 154.7 & 8.12 &  2007JUL22 \\
07.632$-$0.109  &  18 03 11.63  &  -22 24 32.4  & 146.5 & 158.9 & 157.0 & 6.55 & 157.0 & 6.26 &  2007JUL22\\
08.139+0.226  &  18 03 00.75  &  -21 48 09.9  & 18.8 & 21.8 & 19.9 & 11.40 & 19.9 & 5.24 &  C2009\\
08.317$-$0.096  &  18 04 36.02  &  -21 48 19.6  & 44.0 & 49.2 & 47.6 & 3.82 & 47.1 & 3.03 &  2007JUL21 \\
08.669$-$0.356  &  18 06 18.99  &  -21 37 32.2  & 35.8 & 39.7 & 39.2 & 9.96 & 39.0 & 10.46 &  C2009 \\
08.683$-$0.368  &  18 06 23.49  &  -21 37 10.2  & 35.8 & 45.6 & 43.2 & 102.00 & 43.2 & 141.70 &  C2009 \\
08.832$-$0.028  &  18 05 25.67  &  -21 19 25.1  & -6.0 & 5.9 & -3.8 & 159.08 & -3.8 & 126.80 &  2007MAR21; (X09) \\
08.872$-$0.493  &  18 07 15.34  &  -21 30 53.7  & 22.5 & 27.5 & 23.3 & 33.86 & 23.3 & 27.37 &  2007JUL21; (X09) \\
09.215$-$0.202  &  18 06 52.84  &  -21 04 27.5  & 36.0 & 50.0 & 45.5 & 11.96 & 45.6 & 9.16 &  2007JUL21 \\
09.621+0.196  &  18 06 14.67  &  -20 31 32.4  & -4.8 & 8.9 & 1.3 & 5239.85 & 1.3 & 5196.00 &  C2009 \\
09.619+0.193  &  18 06 14.92  &  -20 31 44.3  & 5.0 & 7.0 & 5.5 & 70.00 & 5.4 & 65.00 &  C2009 \\
09.986$-$0.028  &  18 07 50.12  &  -20 18 56.5  & 40.6 & 51.8 & 42.2 & 67.58 & 42.2 & 70.24 &  C2009 \\
10.205$-$0.345  &  18 09 28.43  &  -20 16 42.5  & 5.6 & 11.0 & 6.6 & 2.03 & 7.2 & 1.28 &  2008OCT20 \\
10.287$-$0.125  &  18 08 49.36  &  -20 05 59.0  & 1.5 & 6.0 & 4.6 & 7.19 & 4.5 & 8.29 &  C2009; (W98) \\
10.299$-$0.146  &  18 08 55.54  &  -20 05 57.5  & 19.0 & 21.0 & 19.9 & 0.94 & 20.0 & 1.51 &  C2009 \\
10.320$-$0.259  &  18 09 23.30  &  -20 08 06.9  & 35.0 & 39.6 & 39.0 & 9.50 & 39.0 & 8.87 &  C2009 \\
10.323$-$0.160  &  18 09 01.46  &  -20 05 07.8  & 4.0 & 16.0 & 11.5 & 90.05 & 11.6 & 94.62 &  C2009 \\
10.342$-$0.142  &  18 08 59.99  &  -20 03 35.4  & 6.0 & 18.0 & 15.4 & 15.05 & 14.8 & 14.00 &  C2009 \\
10.356$-$0.148& 18 09 03.07 & -20 03 02.2 & 49.6 & 54.2 & 49.9 & 1.45&50.0& 1.20& 2010MAY22\\
10.444$-$0.018  &  18 08 44.88  &  -19 54 38.2  & 67.6 & 79.0 & 73.3 & 24.27 & 73.4 & 24.92 &  C2009 \\
10.472+0.027  &  18 08 38.20  &  -19 51 50.1  & 57.5 & 77.6 & 75.0 & 28.01 & 75.1 & 35.07 &  C2009 \\ 
10.480+0.033  &  18 08 37.88  &  -19 51 16.1  & 57.0 & 66.0 & 59.5 & 22.53 & 59.5 & 24.07 &  C2009\\
10.627$-$0.384  &  18 10 29.22  &  -19 55 41.1  & -6.0 & 7.7 & 4.6 & 4.23 & 4.6 & 3.78 &  C2009 \\
10.629$-$0.333  &  18 10 17.98  &  -19 54 04.8  & -13.5 & 1.0 & -0.2 & 4.97 & -0.4 & 4.20 &  C2009 \\
10.724$-$0.334  &  18 10 30.03  &  -19 49 06.8  & -2.5 & -1.6 & -2.2 & 4.78 & -2.1 & 1.51 &  2007FEB05 \\
10.822$-$0.103  &  18 09 50.52  &  -19 37 14.1  & 68.0 & 74.0 & 72.0 & 0.28 & 72.1 & 0.90 &  2007FEB06 \\
10.886+0.123  &  18 09 07.98  &  -19 27 21.8  & 14.0 & 22.5 & 17.1 & 12.03 & 17.2 & 12.24 &  2007FEB06 \\
10.958+0.022  &  18 09 39.32  &  -19 26 28.0  & 23.0 & 25.5 & 24.5 & 12.30 & 24.5 & 16.63 &  C2009 \\
11.034+0.062  &  18 09 39.84  &  -19 21 20.3  & 15.2 & 21.0 & 20.5 & 0.55 & 20.6 & 0.89 &  C2009 \\
11.109$-$0.114  &  18 10 28.25  &  -19 22 29.1  & 22.0 & 34.5 & 24.0 & 15.02 & 24.1 & 14.81 &  2007FEB05 \\
11.497$-$1.485  &  18 16 22.13  &  -19 41 27.1  & 4.8 & 17.0 & 6.6 & 68.40 & 6.7 & 61.75 &  C2009 \\
11.903$-$0.102  &  18 12 02.70  &  -18 40 24.7  & 32.0 & 36.7 & 33.9 & 11.30 & 33.8 & 11.60 &  C2009 \\
11.904$-$0.141  &  18 12 11.44  &  -18 41 28.6  & 39.5 & 45.0 & 42.9 & 64.89 & 42.8 & 66.53 &  C2009 \\
11.936$-$0.150  &  18 12 17.29  &  -18 40 02.6  & 46.0 & 50.0 & 48.5 & 2.15 & 48.4 & 2.30 &  C2009 \\
11.936$-$0.616  &  18 14 00.89  &  -18 53 26.6  & 30.1 & 44.6 & 32.3 & 42.91 & 32.2 & 49.32 &  C2009 \\
11.992$-$0.272  &  18 12 51.19  &  -18 40 38.4  & 56.0 & 60.5 & 59.8 & 1.89 & 59.8 & 1.60 &  2007FEB06; (W98) \\
12.025$-$0.031  &  18 12 01.86  &  -18 31 55.7  & 105.0 & 113.1 & 108.3 & 96.26 & 108.3 & 103.10 &  C2009 \\
12.112$-$0.126  &  18 12 33.39  &  -18 30 07.6  & 38.0 & 52.0 & 39.9 & 2.98 & 39.9 & 2.80 &  2008JAN23 \\
12.181$-$0.123  &  18 12 41.00  &  -18 26 21.9  & 29.0 & 31.0 & 29.7 & 1.92 & 29.7 & 1.40 &  C2009 \\
12.199$-$0.033  &  18 12 23.44  &  -18 22 50.7  & 48.2 & 57.1 & 49.3 & 13.67 & 49.3 & 16.91 &  C2009; (2008JAN23) \\
12.202$-$0.120  &  18 12 42.93  &  -18 25 11.8  & 26.0 & 27.0 & 26.4 & 0.80 & 26.4 & 0.60 &  C2009 \\
12.203$-$0.107  &  18 12 40.24  &  -18 24 47.5  & 20.0 & 32.0 & 20.5 & 2.43 & 20.5 & 1.83 &  C2009 \\
12.209$-$0.102  &  18 12 39.92  &  -18 24 17.9  & 16.0 & 22.0 & 19.8 & 11.48 & 19.8 & 10.16 &  C2009 \\
12.265$-$0.051  &  18 12 35.40  &  -18 19 52.3  & 58.0 & 70.9 & 68.5 & 2.25 & 68.3 & 2.56 &  C2009 \\
12.526+0.016  &  18 12 52.04  &  -18 04 13.6  & 38.8 & 44.0 & 42.6 & 3.07 & 42.5 & 3.30 &  2008JAN23 \\
12.625$-$0.017  &  18 13 11.30  &  -17 59 57.6  & 21.2 & 28.0 & 21.6 & 25.49 & 21.6 & 23.61 &  C2009; (B02) \\
12.681$-$0.182  &  18 13 54.75  &  -18 01 46.6  & 50.0 & 62.0 & 57.5 & 350.98 & 57.5 & 350.50 &  C2009 \\
12.776+0.128  &  18 12 57.57  &  -17 47 49.2  & 30.4 & 33.0 & 32.8 & 0.84 & 32.9 & 0.90 &  2007NOV25 \\
12.889+0.489  &  18 11 51.40  &  -17 31 29.6  & 28.0 & 43.0 & 39.3 & 68.88 & 39.2 & 78.93 &  C2009\\
12.904$-$0.031  &  18 13 48.27  &  -17 45 38.8  & 55.8 & 61.0 & 59.1 & 19.99 & 59.1 & 40.34 &  2008JAN23\\
12.909$-$0.260  &  18 14 39.53  &  -17 52 00.0  & 34.7 & 47.0 & 39.9 & 269.05 & 39.9 & 245.48 &  C2009 \\
\hline
\end{tabular} 
\label{resotable}
\end{table*}

\begin{table*} \centering
\addtocounter{table}{-1}
 \caption{\small Positions and parameters of methanol masers cont.} 
\begin{tabular}{lcrcccrcrl}
\hline

\multicolumn{1}{c}{Source Name} & \multicolumn{2}{c}{Equatorial
Coordinates} & \multicolumn{2}{c}{Velocity range} & \multicolumn{2}{c}{MX
data} & \multicolumn{2}{c}{Survey Cube data} & \multicolumn{1}{l}{Position}\\

\ (~~~l,~~~~~~~b~~~)    &       RA(2000)        &       Dec(2000)       &
$\rm V_{L}$&$\rm V_{H}$ &  $\rm V_{pk}$(MX)     &  $\rm S_{pk}$(MX)  &
$\rm V_{pk}$(SC) & $\rm S_{pk}$(SC)     &  Refs,
epoch     \\
\ (~~~$^\circ$~~~~~~~$^\circ$~~~) & (h~~m~~~s) & (~$^\circ$~~ '~~~~") &
\multicolumn{2}{c}{(km\,s$^{-1}$ )} & (km\,s$^{-1}$ ) &  (Jy) & (km\,s$^{-1}$ ) & (Jy) \\
\hline
13.179+0.061  &  18 14 00.96  &  -17 28 32.5  & 45.6 & 50.0 & 46.5 & 1.48 & 46.5 & 2.57 &  2008JAN23 \\
 13.657$-$0.599  &  18 17 24.26  &  -17 22 12.5  & 45.0 & 52.7 & 51.2 & 32.22 & 51.3 & 33.71 &  C2009; (2008JAN23) \\
13.696$-$0.156  &  18 15 51.05  &  -17 07 29.6  & 98.3 & 108.5 & 99.3 & 1.90 & 99.4 & 1.86 &  2007NOV25 \\
13.713$-$0.083  &  18 15 36.99  &  -17 04 31.8  & 43.0 & 53.2 & 43.6 & 12.74 & 43.6 & 12.63 &  2008JAN23; (2007NOV25)\\
14.101+0.087  &  18 15 45.81  &  -16 39 09.4  & 4.4 & 16.6 & 15.4 & 87.26 & 15.4 & 86.55 &  C2009 \\
14.230$-$0.509  &  18 18 12.59  &  -16 49 22.8  & 24.6 & 26.7 & 25.3 & 0.20 & 25.3 & 3.62 &  2009JAN09; (2008OCT20) \\
14.390$-$0.020  &  18 16 43.77  &  -16 27 01.0  & 24.5 & 28.5 & 26.9 & 3.12 & 26.9 & 4.40 &  2008OCT20 \\
14.457$-$0.143  &  18 17 18.79  &  -16 27 57.5  & 38.0 & 44.2 & 43.2 & 0.81 & 43.1 & 1.48 &  2008OCT20 \\
14.490+0.014  &  18 16 48.06  &  -16 20 45.0  & 19.8 & 24.5 & 20.2 & 1.28 & 20.2 & 1.15 &  2008OCT20 \\
14.521+0.155  &  18 16 20.73  &  -16 15 05.5  & -3.0 & 6.0 & 4.1 & 1.40 & 4.1 & 1.31 &  2008OCT20 \\
14.604+0.017  &  18 17 01.14  &  -16 14 38.0  & 22.1 & 35.8 & 24.6 & 2.33 & 24.7 & 2.30 &  2008AUG23; (W98) \\
14.631$-$0.577  &  18 19 15.21  &  -16 30 04.5  & 23.9 & 25.9 & 25.2 & 1.10 & 25.2 & 1.18 &  2008OCT20 \\
14.991$-$0.121  &  18 18 17.32  &  -15 58 08.3  & 44.6 & 54.0 & 46.0 & 7.25 & 46.0 & 6.63 &  2008AUG23 \\
15.034$-$0.677  &  18 20 24.78  &  -16 11 34.6  & 20.0 & 24.0 & 21.3 & 47.48 & 21.3 & 51.59 &  C2009 \\
15.094+0.192  &  18 17 20.82  &  -15 43 46.5  & 22.5 & 26.5 & 25.7 & 13.63 & 25.8 & 18.62 &  2008AUG23 \\
15.607$-$0.255  &  18 19 59.34  &  -15 29 22.8  & 65.1 & 66.5 & 65.9 & 0.43 & 66.0 & 0.85 &  2008OCT20 \\
15.665$-$0.499  &  18 20 59.75  &  -15 33 10.0  & -5.0 & -2.0 & -2.9 & 42.60 & -2.9 & 47.65 &  2008AUG23 \\
16.112$-$0.303  &  18 21 09.14  &  -15 04 00.6  & 33.4 & 35.6 & 34.5 & 2.05 & 34.5 & 2.22 &  2008AUG23 \\
16.302$-$0.196  &  18 21 07.83  &  -14 50 54.6  & 46.9 & 53.6 & 51.9 & 11.24 & 51.8 & 9.91 &  2008AUG23 \\
16.403$-$0.181  &  18 21 16.39  &  -14 45 09.0  & 39.1 & 40.0 & 39.2 & 0.47 & 39.2 & 0.50 &  2008OCT20 \\
16.585$-$0.051  &  18 21 09.13  &  -14 31 48.5  & 52.0 & 69.5 & 62.1 & 36.74 & 62.1 & 33.30 &  2008AUG23; (W98) \\
16.662$-$0.331  &  18 22 19.46  &  -14 35 39.1  & 42.5 & 44.2 & 43.0 & 3.19 & 43.0 & 2.56 &  2008AUG23 \\
16.831+0.079  &  18 21 09.53  &  -14 15 08.6  & 57.2 & 69.4 & 58.7 & 4.06 & 58.7 & 4.44 &  2008AUG23 \\
16.855+0.641  &  18 19 09.57  &  -13 57 57.5  & 23.0 & 25.0 & 24.2 & 1.67 & 24.2 & 1.49 &  2008OCT20 \\
16.864$-$2.159  &  18 29 24.42  &  -15 16 04.5  & 14.0 & 20.0 & 15.0 & 28.93 &  -  &  -  &  2008OCT20; (W98) \\
16.976$-$0.005  &  18 21 44.68  &  -14 09 48.5  & 5.0 & 9.0 & 6.5 & 0.64 & 6.6 & 0.70 &  2008OCT20 \\
17.021$-$2.403  &  18 30 36.30  &  -15 14 28.5  & 17.0 & 25.0 & 23.6 & 4.90 &  -  &  -  &  2008OCT20 \\
17.029$-$0.071  &  18 22 05.21  &  -14 08 51.0  & 90.4 & 96.0 & 91.4 & 1.10 & 91.4 & 1.69 &  2008AUG23 \\
17.638+0.157  &  18 22 26.30  &  -13 30 12.1  & 20.0 & 22.0 & 20.8 & 24.79 & 20.8 & 34.17 &  C2009 \\
17.862+0.074  &  18 23 10.10  &  -13 20 40.8  & 107.3 & 119.6 & 110.6 & 1.44 & 110.5 & 1.42 &  2008AUG23 \\
18.073+0.077  &  18 23 33.98  &  -13 09 25.0  & 44.0 & 57.5 & 55.6 & 5.74 & 55.8 & 6.00 &  2008OCT20 \\
18.159+0.094  &  18 23 40.18  &  -13 04 21.0  & 54.0 & 60.0 & 58.4 & 8.46 & 59.0 & 10.25 &  2008AUG23 \\
18.262$-$0.244  &  18 25 05.70  &  -13 08 23.2  & 72.0 & 81.0 & 74.3 & 22.48 & 74.2 & 22.98 &  2008AUG23 \\
18.341+1.768  &  18 17 58.13  &  -12 07 24.8  & 26.0 & 32.0 & 28.0 & 96.21 & 28.1 & 99.28 &  2008AUG23; (B02) \\
18.440+0.045  &  18 24 23.32  &  -12 50 52.1  & 57.0 & 66.0 & 61.8 & 1.85 & 61.8 & 2.45 &  2008AUG23 \\
18.460$-$0.004  &  18 24 36.34  &  -12 51 08.6  & 46.7 & 50.0 & 49.4 & 24.23 & 49.4 & 25.30 &  2008AUG23; (W98) \\
18.661+0.034  &  18 24 51.10  &  -12 39 22.5  & 76.0 & 83.0 & 79.0 & 8.87 & 79.0 & 8.00 &  2008AUG23; \\
18.667+0.025  &  18 24 53.78  &  -12 39 20.4  & 76.1 & 81.0 & 76.6 & 8.71 & 76.6 & 8.78 &  2008AUG23 \\
18.733$-$0.224  &  18 25 55.53  &  -12 42 48.9  & 44.0 & 49.0 & 45.8 & 2.34 & 45.9 & 2.30 &  2008AUG23; \\
18.735$-$0.227  &  18 25 56.46  &  -12 42 50.0  & 36.3 & 38.5 & 37.9 & 4.71 & 38.2 & 3.59 &  2008AUG23; \\
18.834$-$0.300  &  18 26 23.66  &  -12 39 38.0  & 37.5 & 44.0 & 41.3 & 5.01 & 41.2 & 7.47 &  2008AUG23\\
18.874+0.053  &  18 25 11.34  &  -12 27 36.8  & 37.7 & 40.6 & 38.7 & 12.77 & 38.7 & 9.80 &  2008AUG23\\
18.888$-$0.475  &  18 27 07.85  &  -12 41 35.9  & 53.0 & 57.6 & 56.5 & 5.70 & 56.4 & 5.56 &  2008AUG23 \\
18.999$-$0.239  &  18 26 29.24  &  -12 29 07.1  & 65.0 & 69.8 & 69.4 & 1.05 & 69.4 & 0.90 &  2008AUG23 \\
19.009$-$0.029  &  18 25 44.78  &  -12 22 46.1  & 53.0 & 63.0 & 55.4 & 19.32 & 55.4 & 20.39 &  2008AUG23 \\
19.249+0.267  &  18 25 08.02  &  -12 01 42.2  & 12.5 & 21.0 & 19.9 & 2.32 & 19.8 & 3.35 &  2008AUG23 \\
19.267+0.349  &  18 24 52.38  &  -11 58 28.2  & 12.5 & 17.5 & 16.2 & 5.33 & 16.3 & 3.40 &  2008AUG23 \\
19.365$-$0.030  &  18 26 25.79  &  -12 03 52.0  & 24.0 & 30.0 & 25.3 & 33.83 & 25.3 & 34.16 &  2008AUG23; (W98) \\
19.472+0.170n  &  18 25 54.72  &  -11 52 33.0  & 17.0 & 23.0 & 21.7 & 17.97 & 21.7 & 15.06 &  C2009; (2008AUG23) \\
19.472+0.170  &  18 25 54.49  &  -11 52 36.5  & 12.7 & 17.7 & 13.8 & 3.30 & 13.8 & 1.97 &  C2009\\
19.486+0.151  &  18 26 00.39  &  -11 52 22.6  & 19.0 & 27.5 & 20.6 & 6.04 & 20.9 & 5.00 &  C2009 \\
19.496+0.115  &  18 26 09.16  &  -11 52 51.7  & 120.0 & 122.0 & 121.2 & 7.55 & 121.3 & 5.60 &  C2009 \\
19.609$-$0.234  &  18 27 37.99  &  -11 56 37.6  & 36.0 & 42.0 & 40.2 & 0.65 & 40.2 & 1.00 &  2007JUL21 \\
19.612$-$0.120  &  18 27 13.48  &  -11 53 15.7  & 52.5  & 53.5  & 53.1 & 0.93 & 53.0 & 0.98 &  W98 \\
19.612$-$0.134  &  18 27 16.52  &  -11 53 38.2  & 49.0 & 61.0 & 56.5 & 12.48 & 56.5 & 11.74 &  2007JUL21; (W98) \\
19.614+0.011  &  18 26 45.24  &  -11 49 31.4  & 30.8 & 34.8 & 32.8 & 3.99 & 32.9 & 3.80 &  2008AUG23 \\
19.667+0.114  & 18 26 28.97&  -11 43 48.9 & 13.4 & 17.8 & 14.2 & 2.20 & 16.3 & 1.52 & 2010MAY22 \\
19.701$-$0.267  &  18 27 55.52  &  -11 52 40.3  & 41.5 & 46.5 & 43.9 & 10.72 & 43.8 & 11.72 &  2007JUL21; (W98) \\
19.755$-$0.128  &  18 27 31.66  &  -11 45 55.0  & 115.5 & 124.0 & 123.1 & 3.63 & 123.1 & 2.31 &  2008AUG23 \\
19.884$-$0.534  &  18 29 14.37  &  -11 50 23.0  & 46.0 & 48.0 & 46.8 & 4.74 & 46.8 & 6.83 &  2008AUG23; (B02) \\
\hline
\end{tabular} 
\label{resotableP2}
\end{table*}

\begin{figure*}
 \begin{center}
 \renewcommand{\baselinestretch}{1.1}
\includegraphics[width=16.5cm]{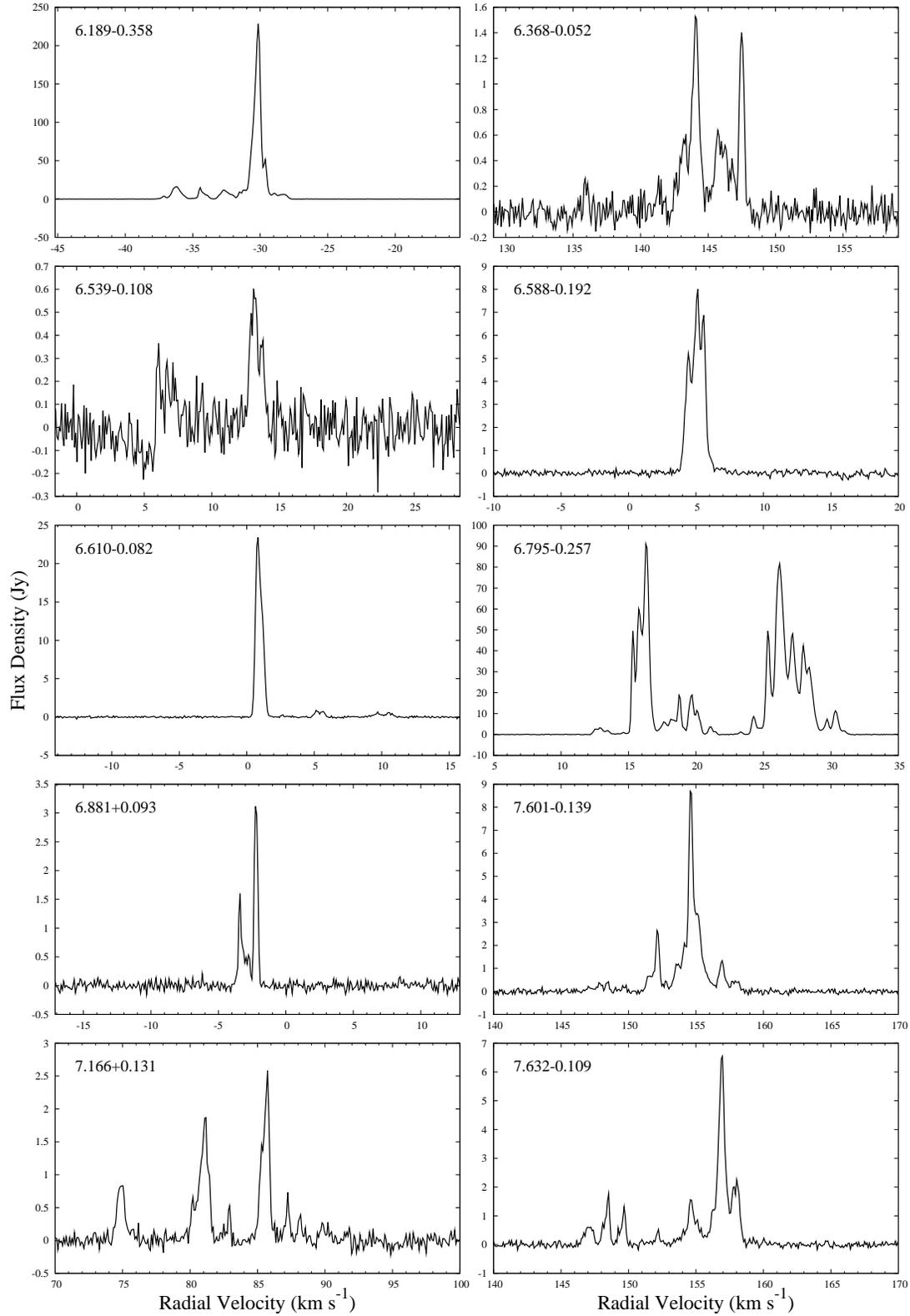}
\caption{\small Spectra of 6668-MHz methanol masers. All spectra are from the MX observations unless stated otherwise.}
\label{spectra}
\end{center}
\end{figure*}

\begin{figure*}
\addtocounter{figure}{-1}
 \begin{center}
 \renewcommand{\baselinestretch}{1.1}
\includegraphics[width=16.5cm]{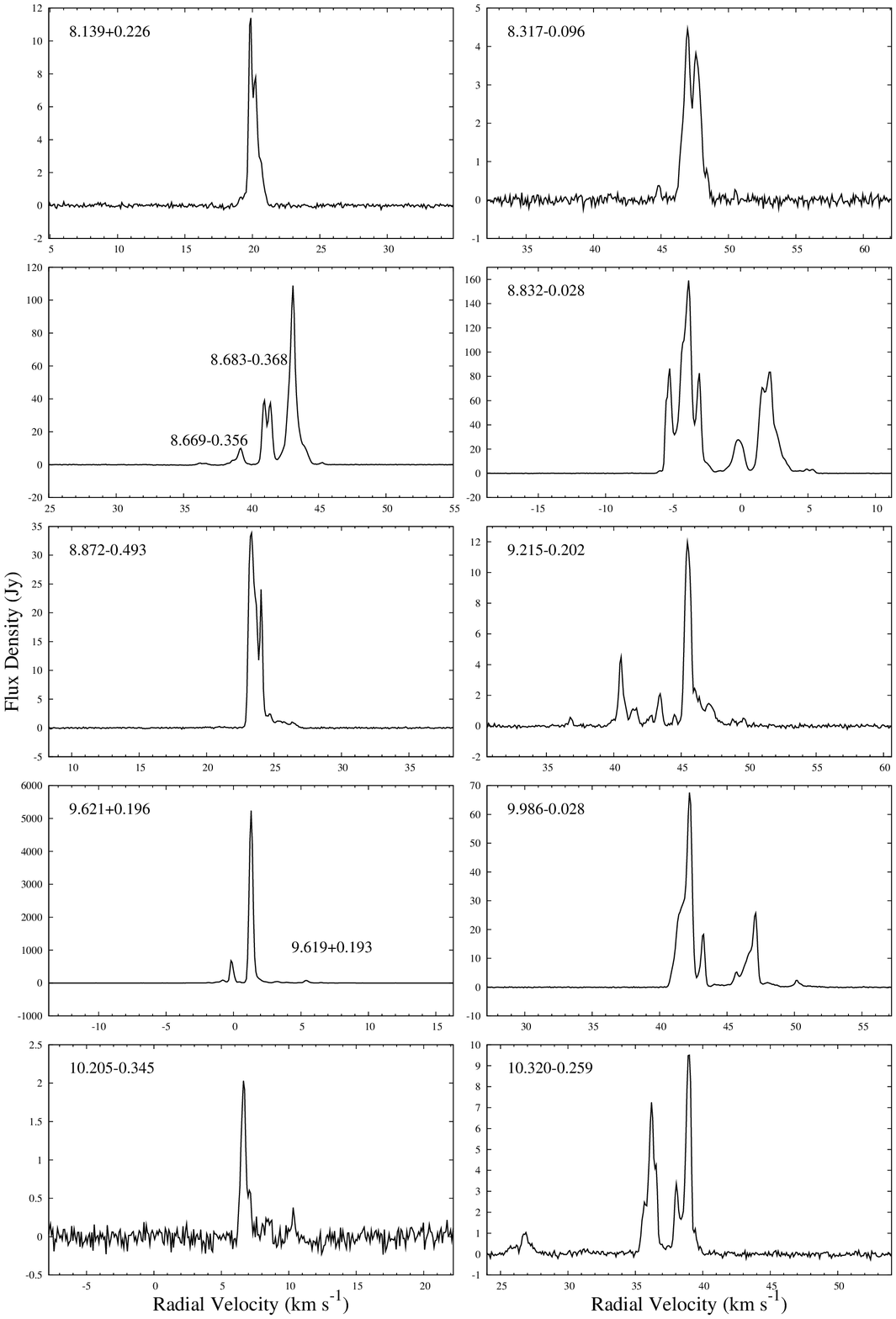}
\caption{\small Spectra of 6668-MHz methanol masers.}
\label{spectra}
\end{center}
\end{figure*}

\begin{figure*}
\addtocounter{figure}{-1}
 \begin{center}
 \renewcommand{\baselinestretch}{1.1}
\includegraphics[width=16.5cm]{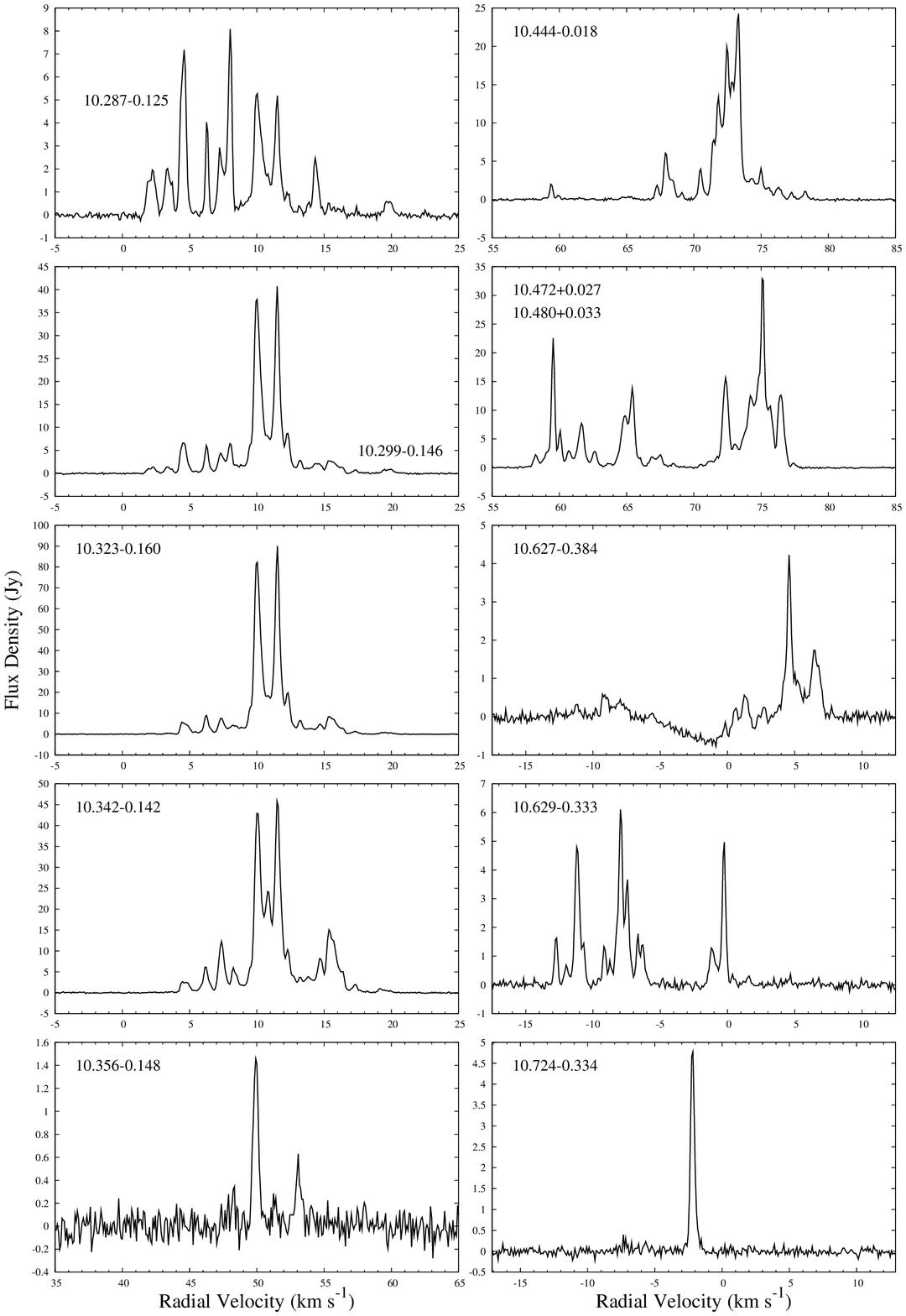}
\caption{\small Spectra of 6668-MHz methanol masers.}
\label{spectra}
\end{center}
\end{figure*}

\begin{figure*}
\addtocounter{figure}{-1}
 \begin{center}
 \renewcommand{\baselinestretch}{1.1}
\includegraphics[width=16.5cm]{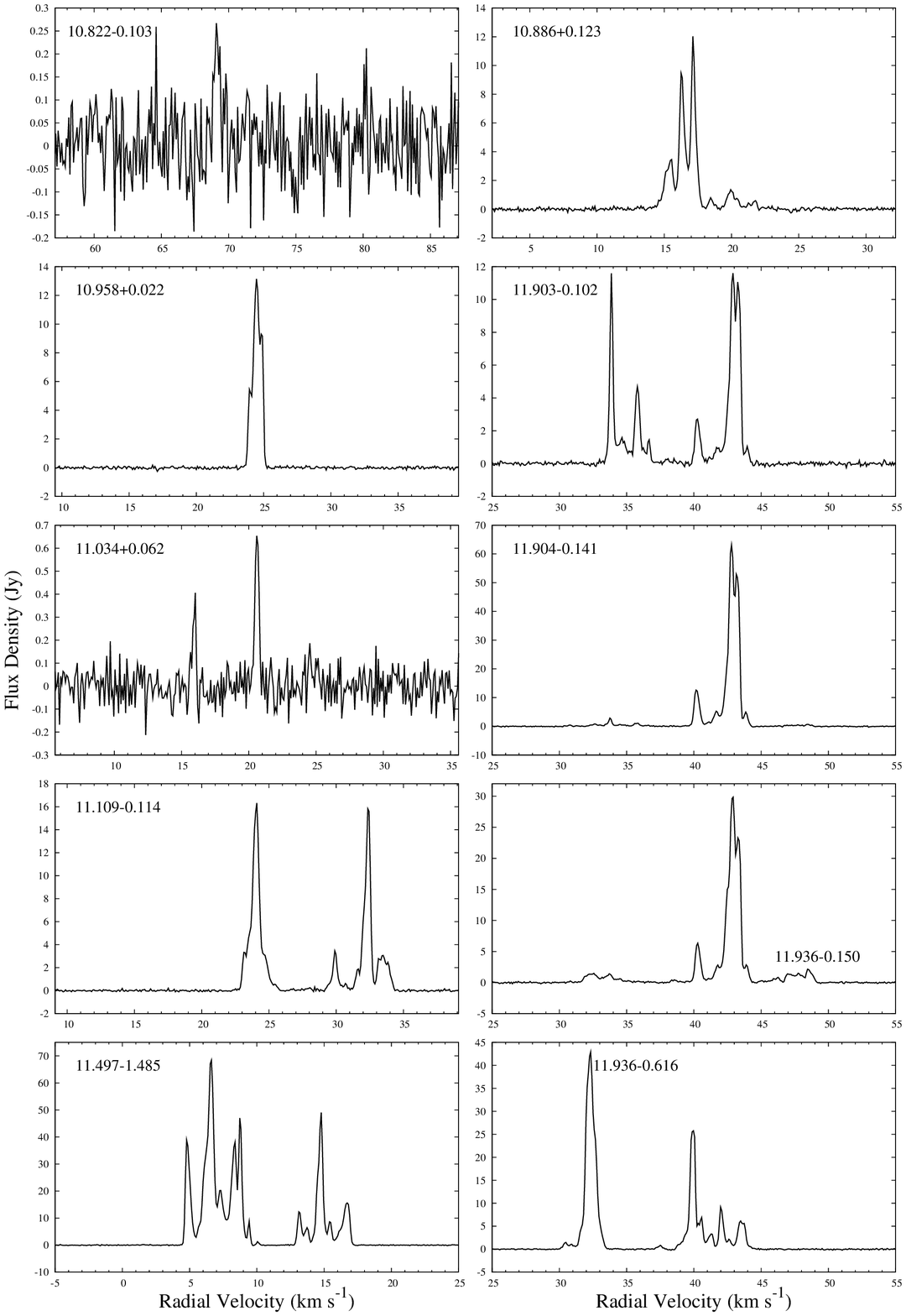}
\caption{\small Spectra of 6668-MHz methanol masers.}
\label{spectra}
\end{center}
\end{figure*}

\begin{figure*}
\addtocounter{figure}{-1}
 \begin{center}
 \renewcommand{\baselinestretch}{1.1}
\includegraphics[width=16.5cm]{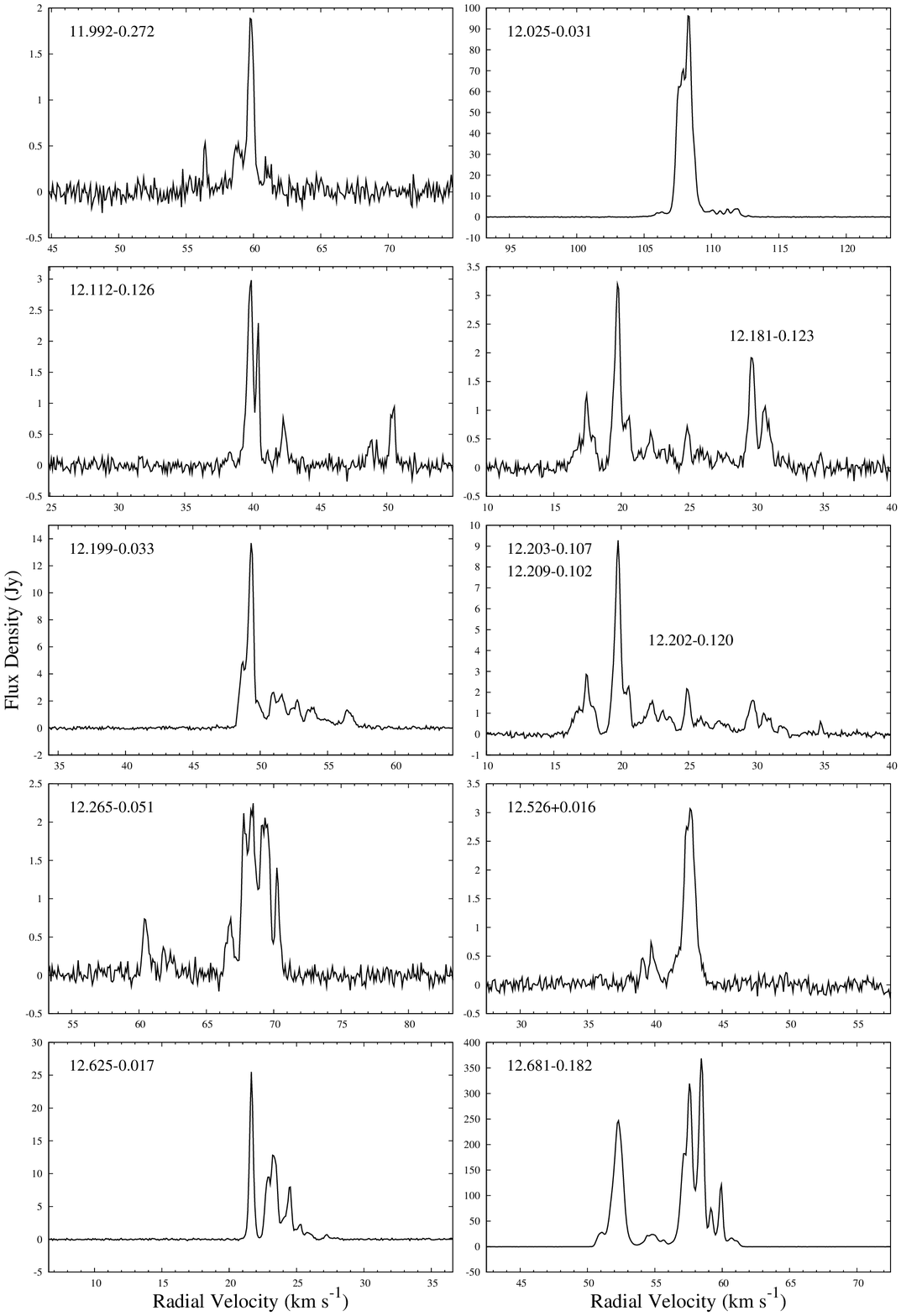}
\caption{\small Spectra of 6668-MHz methanol masers.}
\label{spectra}
\end{center}
\end{figure*}

\begin{figure*}
\addtocounter{figure}{-1}
 \begin{center}
 \renewcommand{\baselinestretch}{1.1}
\includegraphics[width=16.5cm]{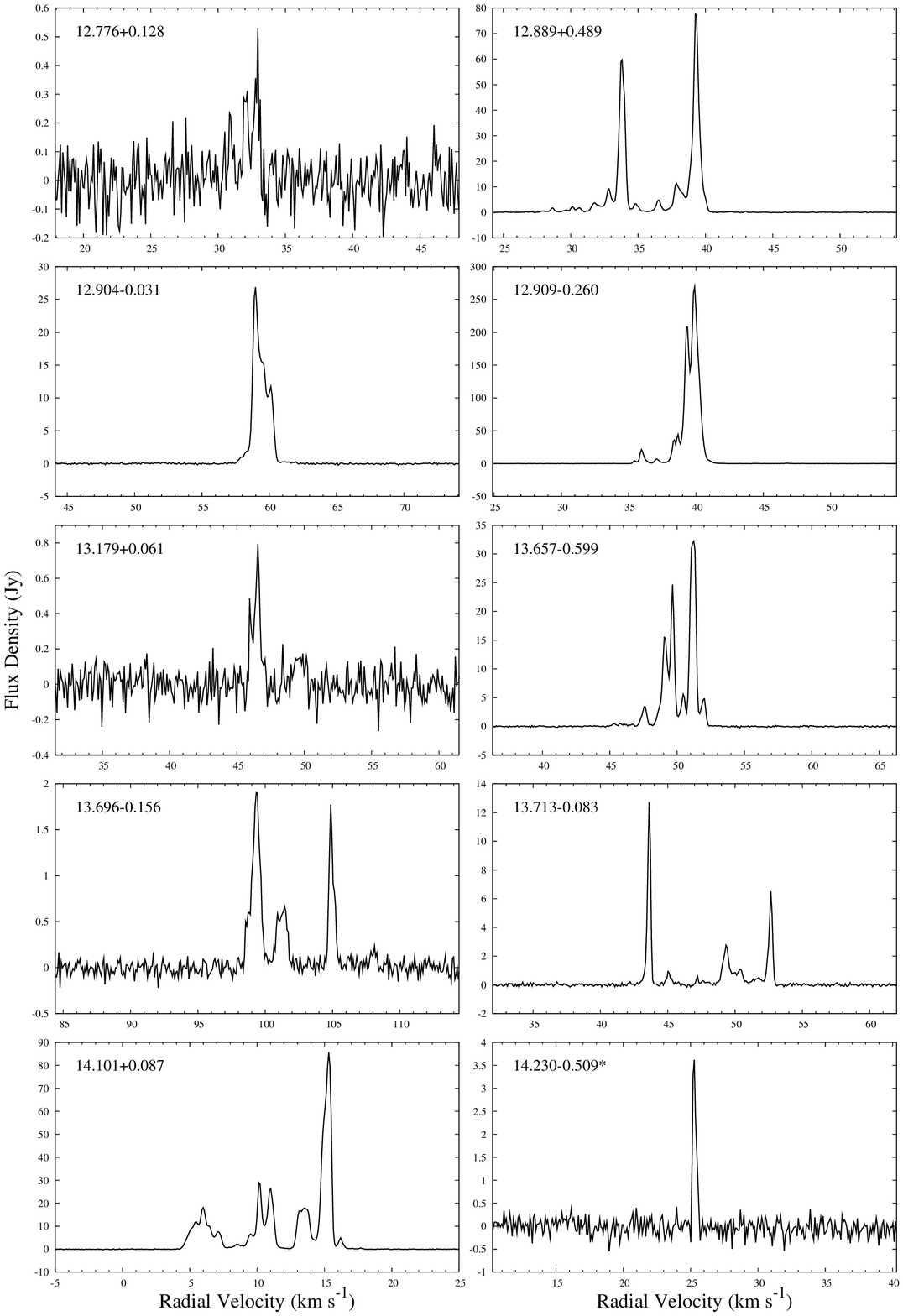}
\caption{\small Spectra of 6668-MHz methanol masers. * denotes survey cube spectrum (source undetected in MX).}
\label{spectra}
\end{center}
\end{figure*}

\begin{figure*}
\addtocounter{figure}{-1}
 \begin{center}
 \renewcommand{\baselinestretch}{1.1}
\includegraphics[width=16.5cm]{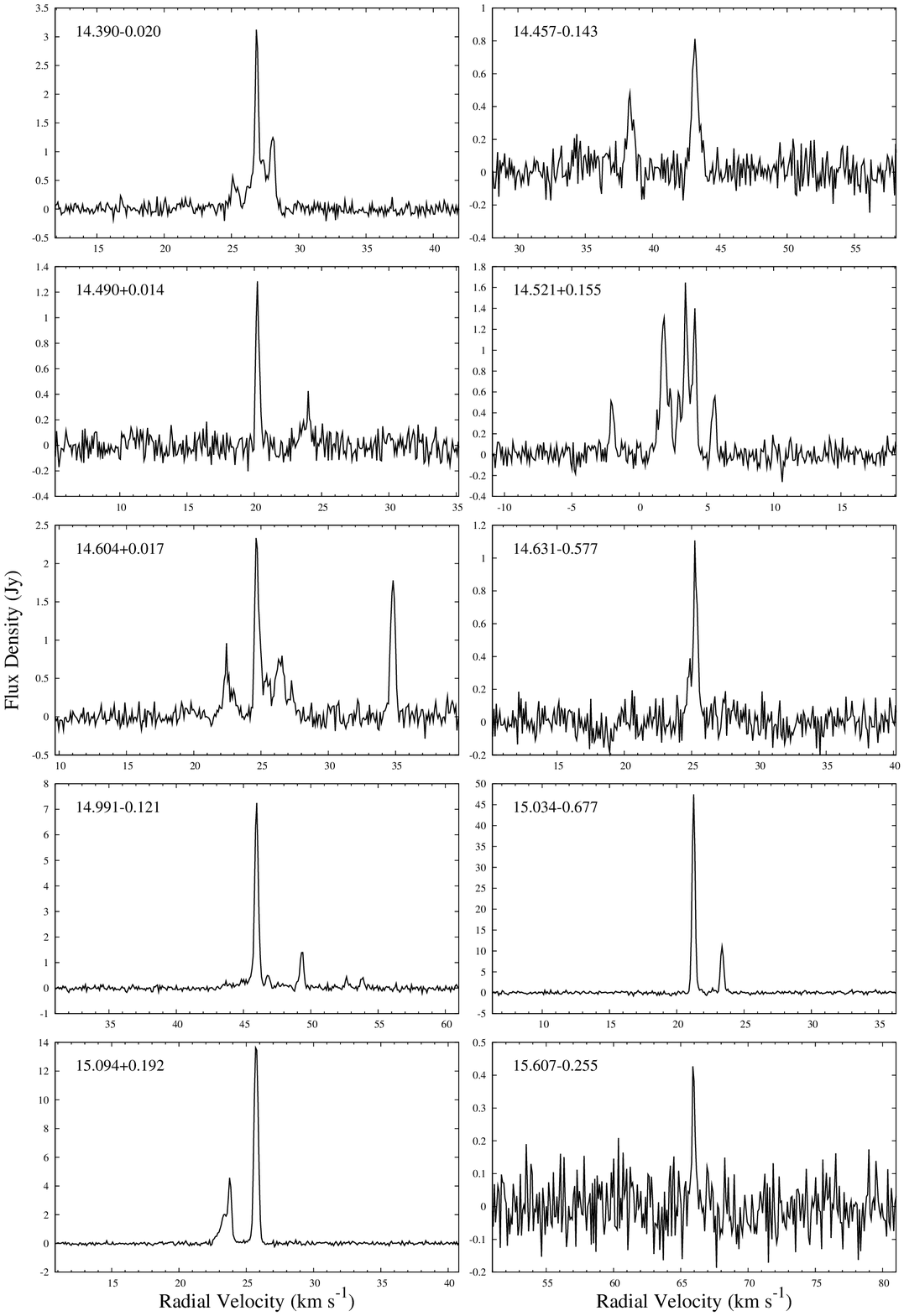}
\caption{\small Spectra of 6668-MHz methanol masers.}
\label{spectra}
\end{center}
\end{figure*}

\begin{figure*}
\addtocounter{figure}{-1}
 \begin{center}
 \renewcommand{\baselinestretch}{1.1}
\includegraphics[width=16.5cm]{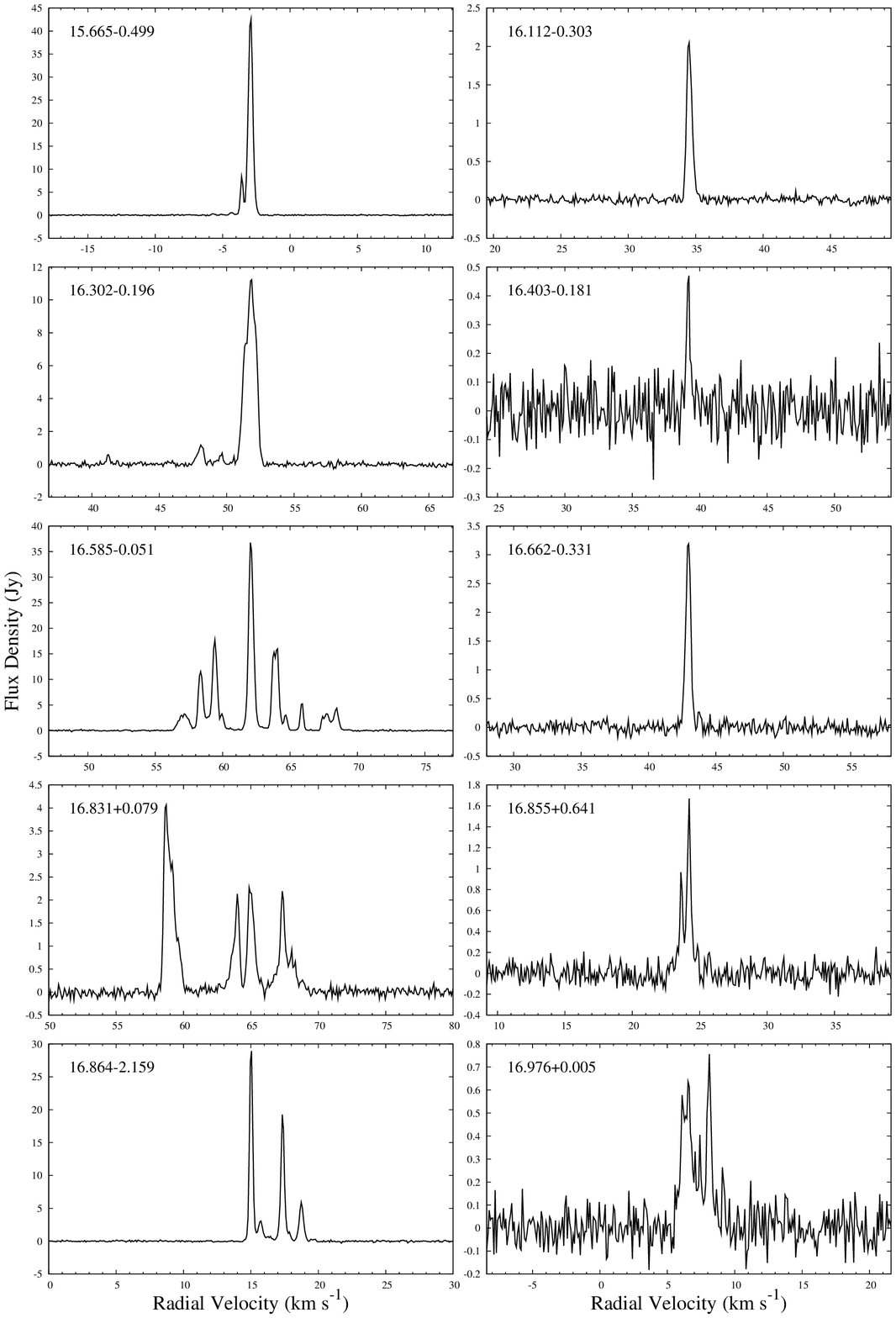}
\caption{\small Spectra of 6668-MHz methanol masers.}
\label{spectra}
\end{center}
\end{figure*}

\begin{figure*}
\addtocounter{figure}{-1}
 \begin{center}
 \renewcommand{\baselinestretch}{1.1}
\includegraphics[width=16.5cm]{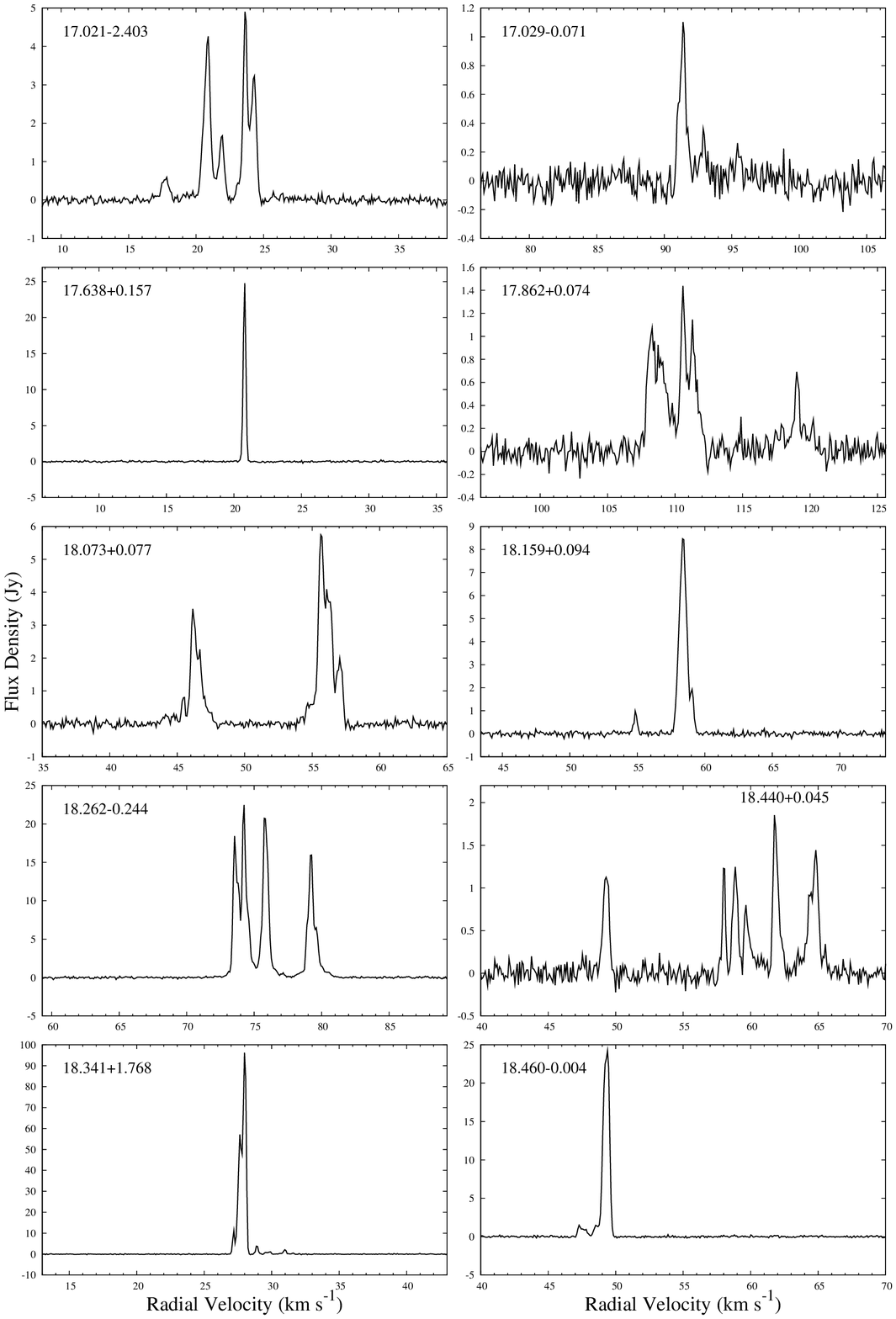}
\caption{\small Spectra of 6668-MHz methanol masers.}
\label{spectra}
\end{center}
\end{figure*}

\begin{figure*}
\addtocounter{figure}{-1}
 \begin{center}
 \renewcommand{\baselinestretch}{1.1}
\includegraphics[width=16.5cm]{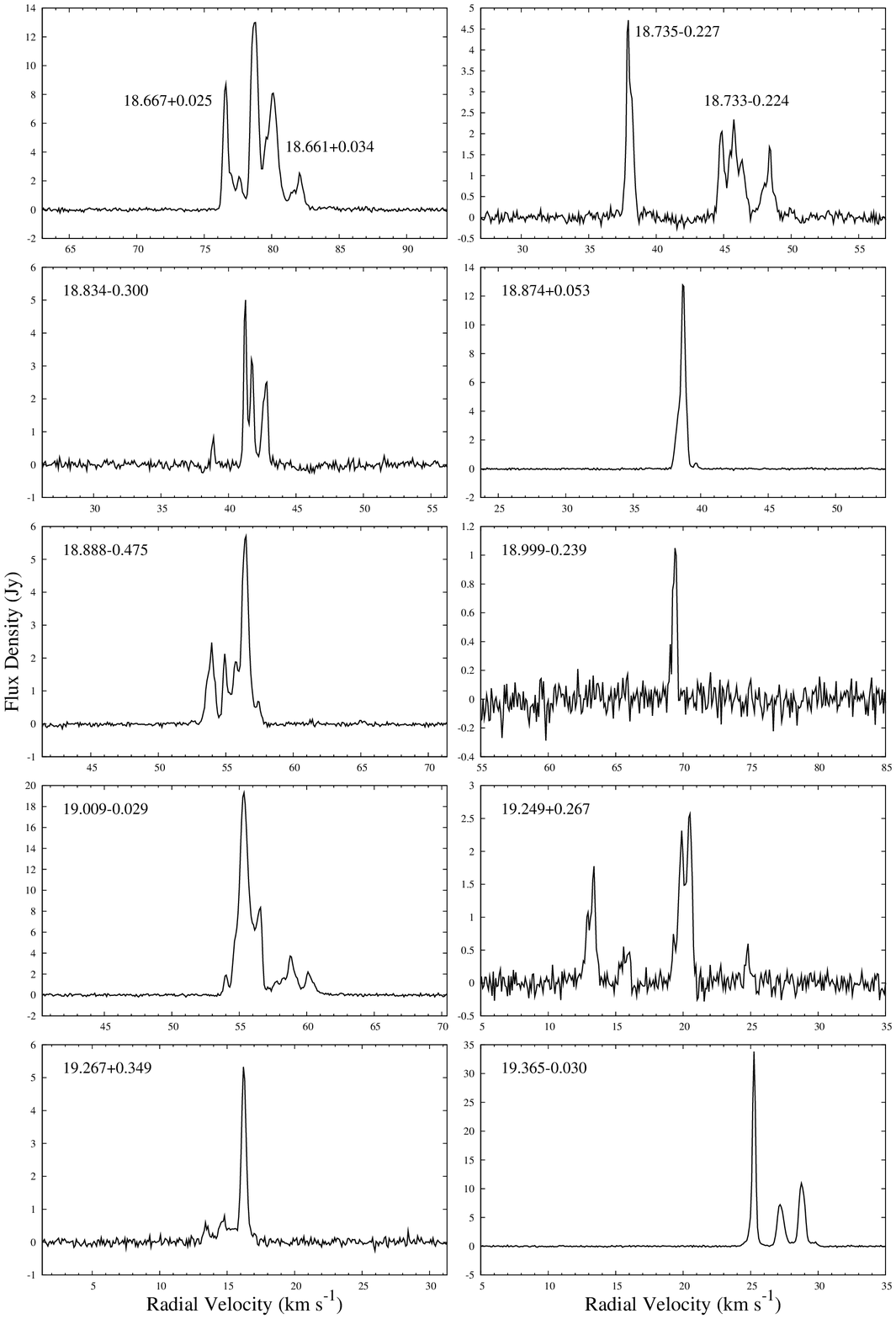}
\caption{\small Spectra of 6668-MHz methanol masers.}
\label{spectra}
\end{center}
\end{figure*}

\begin{figure*}
\addtocounter{figure}{-1}
 \begin{center}
 \renewcommand{\baselinestretch}{1.1}
\includegraphics[width=16.5cm]{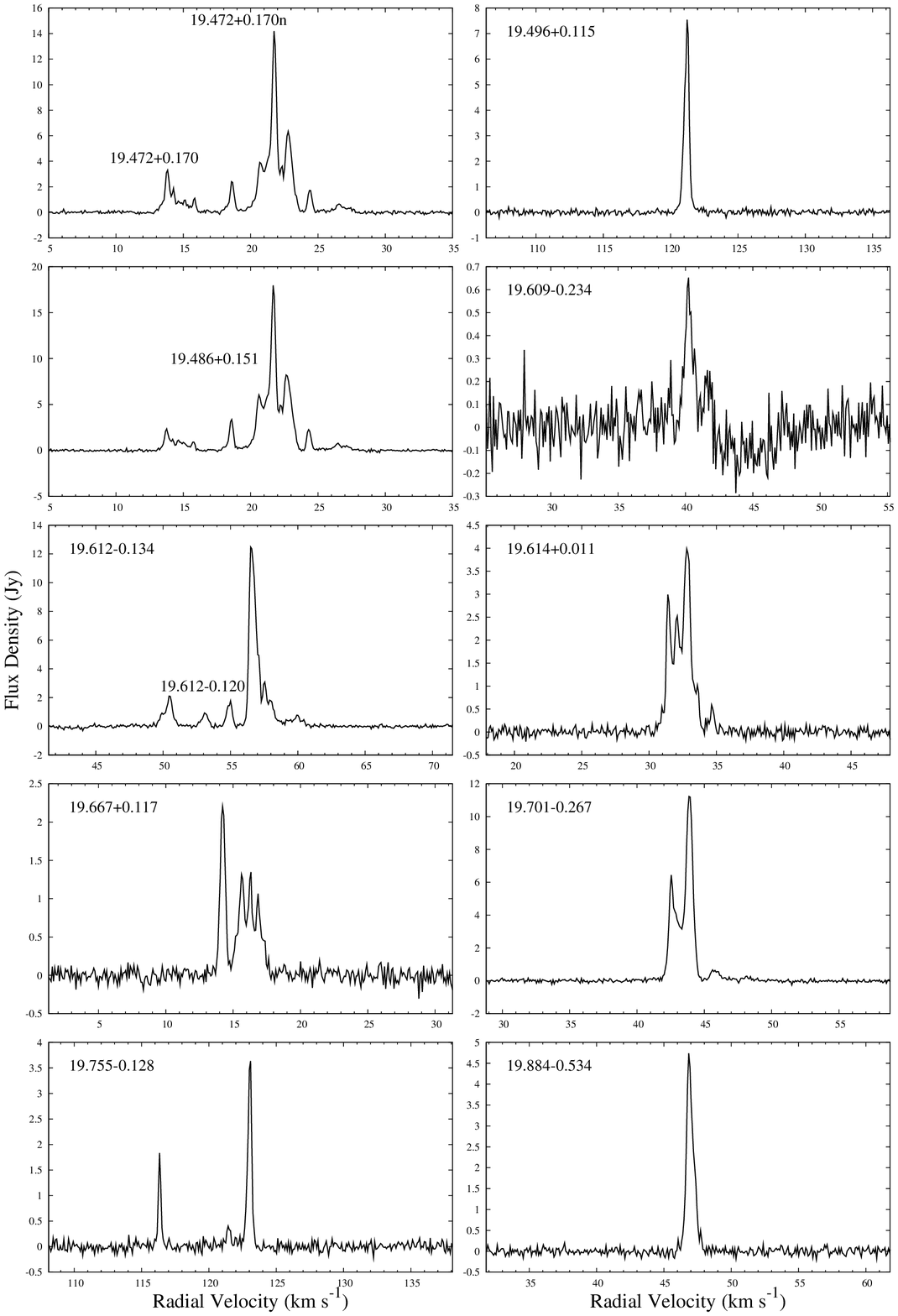}
\caption{\small Spectra of 6668-MHz methanol masers.}
\label{spectra}
\end{center}
\end{figure*}

\section{Discussion} 
Discussion of the global properties of the methanol maser population will be deferred until the full MMB catalogue is published. Here we discuss the properties of the sources within the 6$^{\circ}$ to 20$^{\circ}$ longitude region with reference to those in the Galactic centre region \citep{caswell10mmb1}.

\subsection{Galactic Latitude distribution}
\begin{figure*}
 \begin{center}
 \renewcommand{\baselinestretch}{1.1}
\includegraphics[width=15cm]{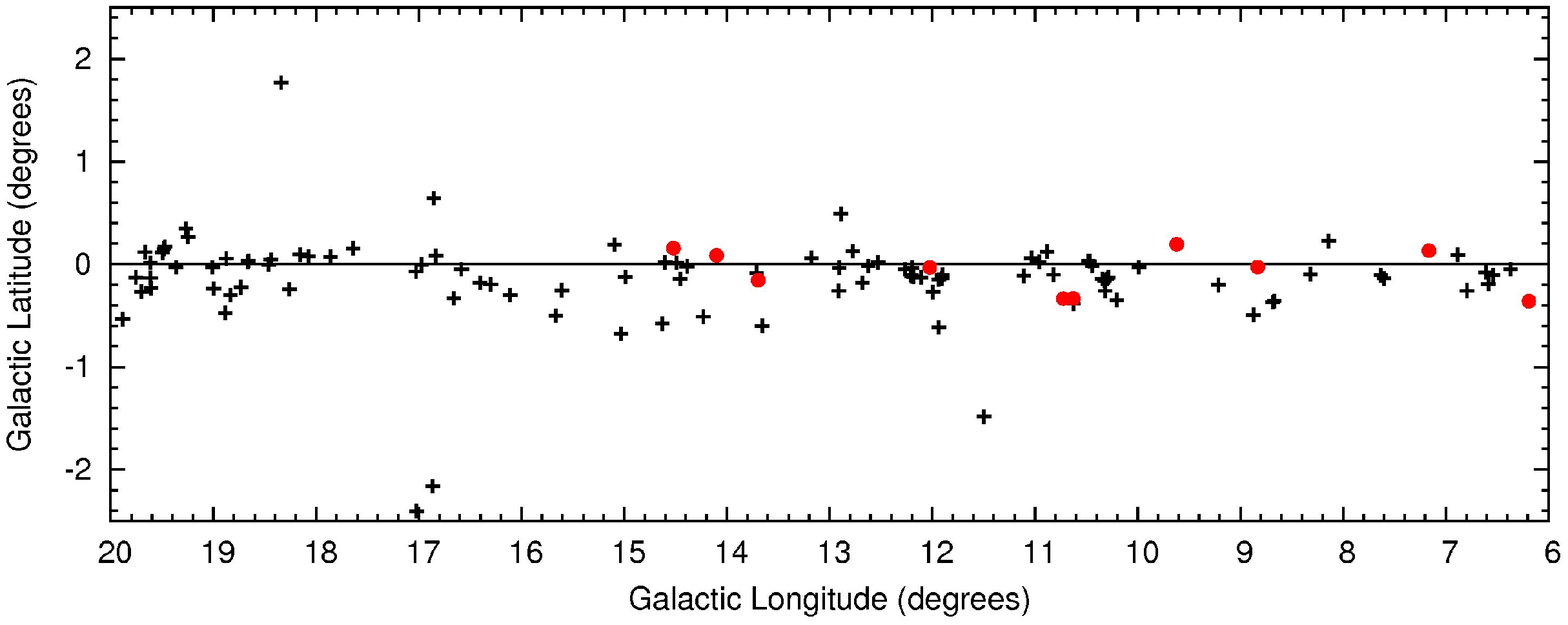}  
\caption{\small Longitude-latitude distribution of sources in the region 6$^{\circ}$ to 20$^{\circ}$. Circles represent sources associated with the 3-kpc arms \citep{green09b}.}
\label{latdist}
\end{center}
\end{figure*}

The maser population in the 6$^\circ$ to 20$^\circ$ longitude region is confined to a narrow range of latitude (Fig. \ref{latdist}); 97\% of the sources (115 out of 119 sources) are at a latitude within 1$^\circ$ of the Galactic plane. ÊThis narrow distribution is very similar to that seen in the 345$^\circ$ to 6$^\circ$ longitude region \citep{caswell10mmb1}. ÊAll four sources with latitudes outside 1$^\circ$ of the plane were previously known (11.497-1.485, 16.864-2.159, 17.021-2.403, 18.341+1.768).

\subsection{Flux densities}
The brightest 6668-MHz methanol maser detected to date, 9.621$+$0.196, is within the current region and was found to have a survey cube peak flux density of $\sim$5200 Jy. The brightest new source detected in the survey is 6.189-0.358 with a survey cube peak flux density of $\sim$220 Jy (and an MX flux density of $\sim$230 Jy). The weakest known source in this region is 6.539-0.108 with a survey cube peak flux density of 0.5 Jy (and an MX flux density of 0.6 Jy). The weakest new detection in the region is 16.983+0.000 with a survey cube peak flux density of 0.7 Jy (0.64 Jy in the follow-up MX observation). Comparable to the 345$^{\circ}$ to 6$^{\circ}$ region we had only three new sources with peak flux densities above 20 Jy: 6.189-0.358, 8.832-0.028 and 8.872-0.493, with survey cube flux densities of 222 Jy, 127 Jy and 27 Jy respectively. There were four sources at or below the survey cube 4 sigma sensitivity limit of 0.7 Jy, two new (16.403-0.181 and 16.976-0.005) and two known (6.539-0.108 and 12.202-0.120). All four sources were confirmed with the higher sensitivity MX observations and ATCA positioning observations.

\subsection{Variability}\label{varisection}
\begin{figure}
 \begin{center}
 \renewcommand{\baselinestretch}{1.1}
\includegraphics[width=7cm]{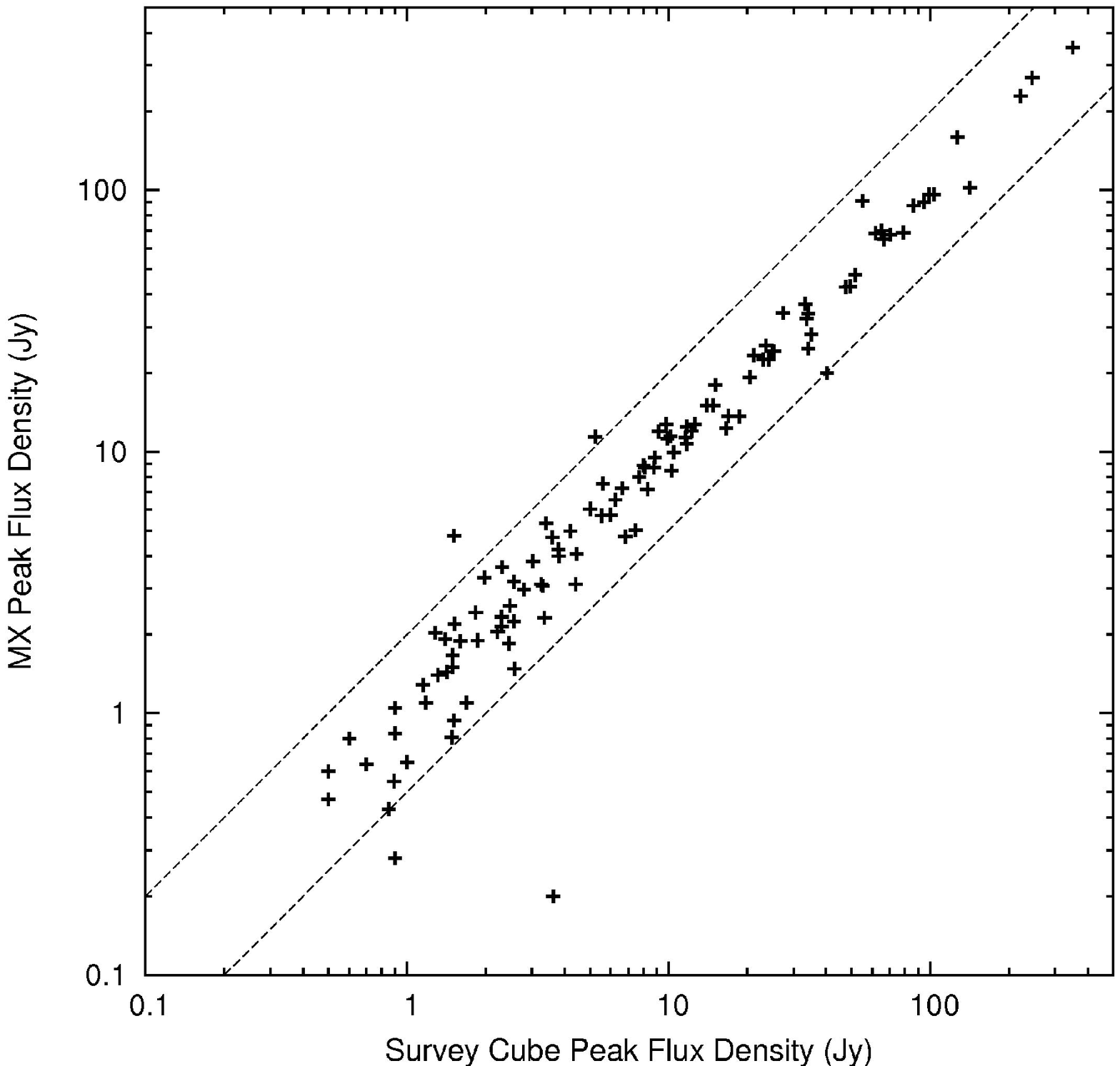}
\caption{\small Variability in source peak flux density between the initial survey observations (Survey Cube) and the later targeted observations (MX) for 6$^{\circ}$  to 20$^{\circ}$  longitude. The dashed lines delimit a factor of 2 variability. The wider spread for weaker sources is likely due to the larger effect of noise variations on the survey cube spectra at these lower flux densities.}
\label{fluxdist}
\end{center}
\end{figure}

The ratio of peak flux density between the survey cube and the MX observations has a median value of 1.01.
Only six sources varied by a factor of 2 or greater, two sources increasing, four decreasing (Fig. \ref{fluxdist}). The two that increased were: 8.139+0.226, which had an MX flux density twice that of the survey cube flux density; and 10.724-0.334, which had an MX peak flux density three times the survey cube value. The four sources which significantly decreased consisted of two with reduced emission and two which disappeared below the detection threshold of the survey. Both 10.822-0.103 and 12.904-0.031 decreased their emission by a factor of $\ge$2 (0.9 to 0.3 Jy and 40 to 20 Jy respectively). The two sources which dropped below our detection threshold represent ideal candidates for variability studies. The first was the new source 14.230-0.509, which had four epochs of observations: it was seen in the survey cube (2007 August) at 3.62 Jy (3 channels $>$ 2 Jy) then subsequently not seen with the ATCA in 2008 October, before being seen again in 2009 January with the ATCA at 1.4 Jy and then not seen again in 2009 March in an MX observation with Parkes. The second was the new source 15.607-0.255, also with four epochs of observations: it was seen in the survey cube data (2007 August) at 0.85 Jy, then reduced to 0.43 Jy in the 2008 March MX before being undetectable in the 2008 August MX (i.e. $<$0.2 Jy) then rising again to $\sim$0.4 Jy in the ATCA observation in 2008 October.
 
In addition to those already mentioned, a further seven sources within longitudes 6$^{\circ}$  to 20$^{\circ}$  are known to be variable (and are individually noted in Section \ref{srcnotes}). Of special note 9.621+0.196 has been shown to periodically flare every 244 days \citep{goedhart04, walt09}, with possible coincident variability in the magnetic field \citep{vlemmings09}. Our survey cube observations were taken 2007 July 8, MJD 54289.5, lying within the rise of the flare \citep{walt09} with the maser having a peak flux density of $\sim$5200 Jy. The MX observation was taken 2008 March 17, MJD 54542.5, lying  within the rise of the subsequent flare, i.e. separated by approximately a full period from the original observation, again with a peak flux density of $\sim$5200 Jy. 12.889+0.489 has also been shown to have a periodic flare, with a period of 29.5 days \citep{goedhart04, goedhart09}. Our survey cube observation was made 2007 August 26, MJD 54338.5, measuring a peak flux density of 79 Jy for the 39.2\,km\,s$^{-1}$ feature. Assuming the 29.5 day period and the time series of \citet{goedhart09}, our survey cube observation is likely to lie in the declining part of a flare. Our MX observation was made 2008 March 17, MJD 54917.5, measuring a peak flux density of 69 Jy for the same feature. According to the time series of \citet{goedhart09} this would place it on the rise of a flare.

\subsection{Individual source velocity spreads}%HERE!
Figure \ref{spectra} shows that the spectra of the masers comprise many narrow spectral features
with individual widths often as small as $\sim$0.3\,km\,s$^{-1}$, spread over a range of velocities.
The total velocity spread in an individual source is dependent on the sensitivity of the observation and can also change as a result of intrinsic variability of the components (Section\,\ref{varisection}).
Therefore, the velocity ranges listed in Table\,\ref{resotable} are the widest velocity spread over which emission has been detected, considering all the observations
available. ÊThis is the same convention as used for the masers in the Galactic
centre \citep{caswell10mmb1}. The mean spread in velocity for a source is 7.2\,km\,s$^{-1}$, the median is 6.0\,km\,s$^{-1}$, and 70\% of the sources have emission spread over velocity ranges less than 10 \,km\,s$^{-1}$. Only three sources of the 119 ($<$3\%) have ranges greater than 16\,km\,s$^{-1}$: 6.795-0.257 (19.3\,km\,s$^{-1}$), 7.166+0.131 (16.5\,km\,s$^{-1}$) and 10.472+0.027 (20.1\,km\,s$^{-1}$). This rarity in velocity spreads greater than 16\,km\,s$^{-1}$ is comparable to both \citet{caswell09a} ($<$3\%) and \citet{caswell10mmb1} ($\sim$1\%).

\subsection{Distribution of velocities} 
Consistent with the results for the central part of the Galaxy \citep{caswell10mmb1}, all the masers in the longitude range 6$^{\circ}$  to 20$^{\circ}$ lie well inside the velocity coverage of the MMB (which was chosen to fully sample the velocity range of the CO emission of \citealt{dame01}). There are three sources in the region with velocities of peak emission exceeding 125\,km\,s$^{-1}$ (6.368-0.052, 7.601-0.139 and 7.632-0.109) and there are six sources with negative velocities of peak emission (6.189-0.358, 6.881+0.093, 8.832-0.028, 10.629-0.333, 10.724-0.334 and 15.665-0.499). Only one of the six sources has a significantly negative velocity ($<$-10 km\,s$^{-1}$), 6.189-0.358. 

%%%%%%%%%%%%%%%%%%%%%%%%%%%%%%%%%%%%%%%%%%%%%%%%%%%%%%%%%%%%

\subsection{Tracing Galactic structure}\label{galstructurediscussion}
6668-MHz methanol masers have the potential to trace the kinematics of Galactic structure: their velocities are closely linked to the systemic velocities \citep{szymczak07, pandian09}; and they are only detected towards regions of high-mass star formation \citep{pestalozzi02b,minier03,xu08}, regions which are intrinsically associated with the spiral arms. Figure \ref{3-kpcExtCir} shows longitudes and velocities of the 6668-MHz methanol sources detected with the MMB survey between longitudes 345$^{\circ}$ and 20$^{\circ}$ (this paper and \citealt{caswell10mmb1}). The masers are shown in relation to the 3-kpc arms as defined by \citet{dame08} and the spiral arm loci based on the logarithmic spirals of \citet{georgelin76}, with the updates of \citet{taylor93} and the rotation curve of \citet{brand93} with IAU standard LSR parameters. Figure \ref{3-kpcExtCir} includes an inset of the assumed spiral arm pattern, oriented so as to readily recognise how individual portions of the arms transform to the {\it l,v} domain. By plotting in the {\it l,v} domain we can investigate Galactic structure without the need to correctly assign individual sources to near and far kinematic distances. The figure shows that there appear to be masers lying between the arms, but before closer inspection of individual anomalies, there are two issues that one must be aware of: the thickness of the arm loci and the choice of rotation curve. Firstly, the arms plotted in {\it l,v} diagrams are purposely thin to reduce obscuration, but as such they do not account for the real spatial radial thickness and velocity dispersion within the arms. 
Secondly, the locations (physical and in the {\it l,v} domain) of the arms depends on the Galactic rotation curve adopted. If the spiral pattern is broadly correct, any change in adopted rotation curve will only alter the position of the arms slightly, leaving them as continuous structures, not only in the Galactic disk, but also in the {\it l,v} domain.  For the Galactic structure investigation  presented here we use the \citet{brand93} curve, since it closely approximates the assumptions made in deriving the spatial pattern of the spiral arms.  Furthermore we resist adopting the most recently suggested rotation curve and revised LSR suggested by \citet{reid09}, as it is preliminary and requires revision \citep{mcmillan10}.  However, we note that near the Galactic Centre, where we did choose the Reid et al. parameters \citep{caswell10mmb1}, this choice had negligible impact on the interpretation of that region. It would affect the current discussion in only one instance (noted below), but would have more impact at longitudes further from the Galactic centre.

\subsubsection{The 3-kpc arm (expanding ring)}\label{3-kpcext}
\begin{figure*}
 \begin{center}
 \renewcommand{\baselinestretch}{1.1}
\includegraphics[width=17cm]{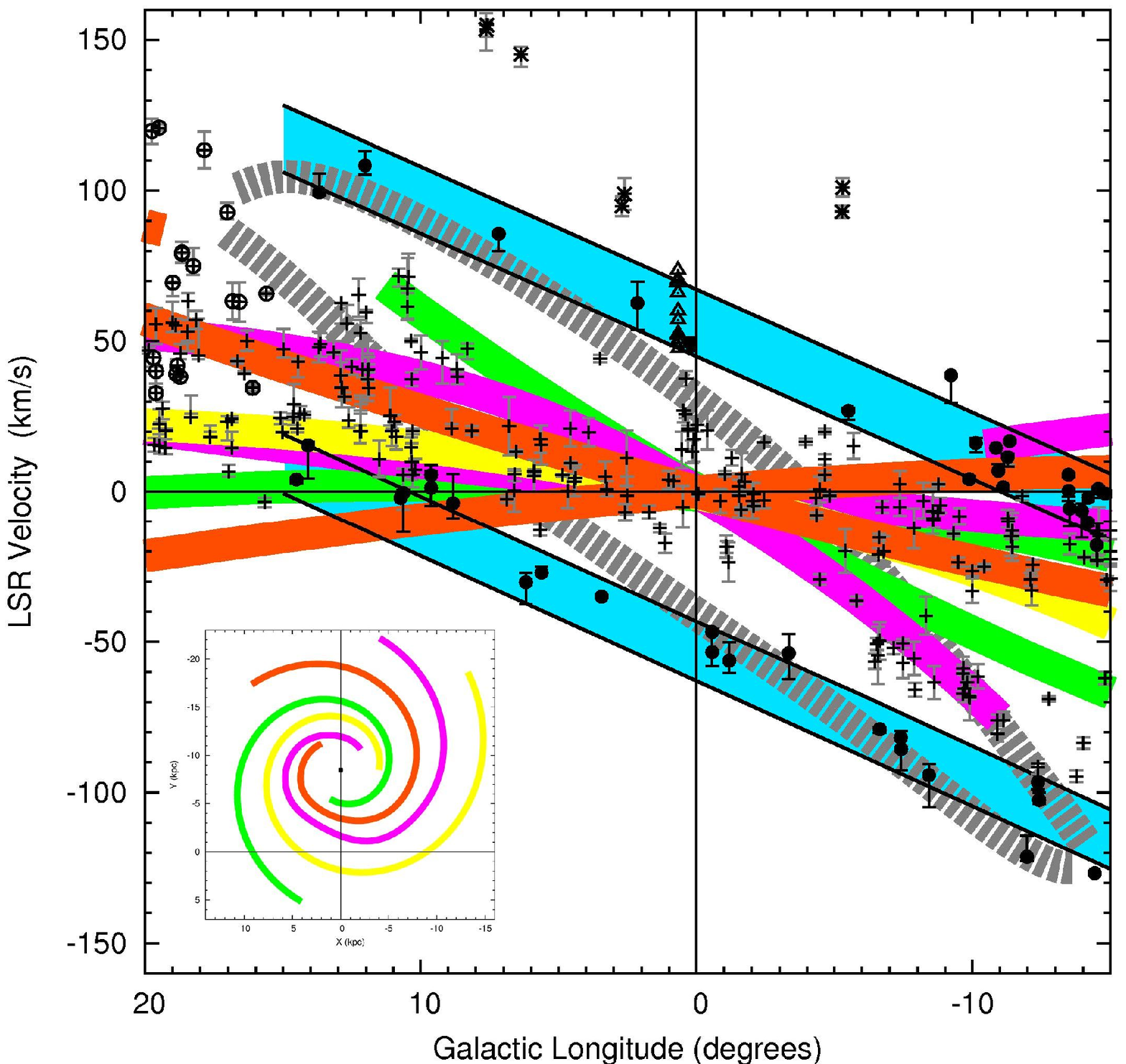}
\caption{\small Longitude-Velocity `crayon' plot showing Methanol Multibeam sources from the current paper and from \citet{caswell10mmb1}. The coloured loci are the spiral arms (Yellow - Perseus; Purple - Carina-Sagittarius; Orange - Crux-Scutum; Green - Norma). The blue shading is the region identified in CO emission as the 3-kpc arms in \citet{dame08}. Shown in broken grey is the locus of an elliptical interpretation of the 3-kpc arm structure \citep{rodriguez08}. Masers are represented by crosses, with the exception of those associated with the 3-kpc arms (filled circles), Sagittarius B2 (open triangles) and the Galactic bar (stars). Crosses surrounded by open circles represent the unassociated sources discussed in Section \ref{3-kpcext}. Grey error bars show the spread over velocity for maser emission. Inlay shows location of arms with respect to the sun (0,0), oriented with positive longitudes to the left in line with the {\it l,v} plot.}
\label{3-kpcExtCir}
\end{center}
\end{figure*}

Table \ref{3-kpclist} lists 45 sources associated kinematically and spatially with the 3-kpc arm features (denoted by filled circles in Figure\,\ref{3-kpcExtCir}). This includes the 42 originally identified in \citet{green09b} and three additional sources, from \citet{caswell10mmb1} and the current paper. A further three sources (two new to the survey) between longitudes 15$^{\circ}$ and 20$^{\circ}$ (17.862+0.074, 19.496+0.115 and 19.755-0.128) have velocities in excess of 100\,km\,s$^{-1}$ and are potential candidates of an extended 3-kpc structure.  They do not align with spiral arm loci (subject to the caveats already mentioned), but they could be accounted for by the far side of the 3-kpc arms if this is extended as a continuous ring structure with a radius of $\sim$3.5\,kpc \citep[e.g.][]{sevenster99}.  An alternative explanation requiring a shift in velocity of the Crux-Scutum arm (by $\sim$30\,km\,s$^{-1}$)  seems less likely. Also within the 15$^{\circ}$ and 20$^{\circ}$ longitude range are eight sources with velocities between 60\,km\,s$^{-1}$ and 100\,km\,s$^{-1}$ (15.607-0.255, 16.585-0.051,16.831+0.079, 17.029-0.071, 18.262-0.244, 18.661+0.034,18.667+0.025 and 18.999-0.239) and seven sources with velocities between 35\,km\,s$^{-1}$ and 50\,km\,s$^{-1}$ (16.112-0.303, 18.735-0.227, 18.834-0.300, 18.874+0.053, 19.609-0.234, 19.614+0.011 and 19.701-0.267) which do not align with spiral arm loci. These sources may also be accounted for by the continuation beyond 15$^{\circ}$ of the 3-kpc arms structure as a ring.  We note that if an alternative rotation model were adopted, the 
shift of the spiral arm loci might account for some of these sources, but would leave a different group of orphans.  For example, using the IAU solar parameter values and flat rotation curve of \citet{reid09} would shift the loci of the Crux-Scutum and Carina-Sagittarius arms by about +15\,km\,s$^{-1}$. This allows 16.585-0.051 (62.1\,km\,s$^{-1}$), 16.831+0.079 (58.7\,km\,s$^{-1}$), 18.999-0.239 (69.4\,km\,s$^{-1}$) and 19.614+0.011 (32.8\,km\,s$^{-1}$)  to be associated with the arms, but disassociates 16.403-0.181 (peak velocity of 39\,km\,s$^{-1}$), 16.662-0.331 (peak velocity of 43\,km\,s$^{-1}$), 18.733-0.224 (peak velocity of 46\,km\,s$^{-1}$) and 19.884-0.534 (peak velocity of 47\,km\,s$^{-1}$). Hence, regardless of parameter and model choice, there are 18 sources unaccounted for by the spiral arms which are candidates for an extended 3-kpc arm structure.

The actual structure of the 3-kpc arms has been variously interpreted as an expanding ring (requiring an explosive precursor), as material orbiting a bar (either on a circular orbit or an elliptical orbit with its major axis parallel to the bar) and as (lateral) arms flowing from the ends of the bar. These physical interpretations attempt to explain the structure seen in longitude-velocity space, most prominently the significant offset from 0\,km\,s$^{-1}$ at longitude 0$^{\circ}$. This anomalous velocity either requires a circular orbit with a radial expansion of $\sim$50\,km\,s$^{-1}$ or an elliptical orbit with appropriate semi-major and minor axes and angle of inclination.
Overall, the results from the MMB so far are in accord with the initial notion of the 3-kpc arms as simply an approximately circular ring uniformly expanding \citep{kruit71}.  More elegant physical descriptions, such as elliptical orbits, appear to have difficulty accounting for the observations.  As an example, we cite a recent suggestion by \citet{rodriguez08}. 
The l-v plot corresponding to their orbits is shown in grey in Figure \ref{3-kpcExtCir}. This model is based on a gas flow model of the 2MASS star count with a disk, bulge and nuclear bar and  comprises arms which extend around the bar forming an ellipse. In this particular example the negative longitudes of the near arm and the positive longitudes of the far arm match both the CO emission (light blue shading in Figure \ref{3-kpcExtCir}) and the 3-kpc maser distribution well. However, for the far arm at negative longitudes the model deviates significantly from the CO emission (although in doing so it does align with a number of masers at approximately $-$10$^{\circ}$ and velocities between $-$50\,km\,s$^{-1}$ and $-$100\,km\,s$^{-1}$). Similarly, for the near arm at positive longitudes, the model also deviates significantly from the CO (although it is possible that modifying the bar characteristics within the model may account for the higher maser density seen between longitudes 10$^{\circ}$ and 15$^{\circ}$). We are currently incorporating MMB data from beyond the longitude range 345$^{\circ}$ to 20$^{\circ}$ into a more comprehensive analysis of the 3-kpc arms, which will be described fully in a later publication.

\subsubsection{The Galactic Bar}
Figure \ref{3-kpcExtCir} highlights seven sources with velocities greater than those of the far 3-kpc arm. Four are previously known and have been associated with the Galactic bar (see \citealt{caswell10mmb1} for details). The other three are newly detected sources between longitudes 6$^{\circ}$ and 7$^{\circ}$. These three sources (6.368-0.052, 7.601-0.139, 7.632-0.109) are also likely to be associated with the bar, having kinematics which fit the near-side of orbits following a barred potential \citep[e.g.][]{binney91, fux99}.
The bar is now traced by methanol masers over the longitude range $-$5.3$^{\circ}$ to $+$7.6$^{\circ}$. 
Emission from an H{\sc ii} region provides evidence that star formation in the bar can also be traced at large negative velocities exceeding $-$200\,km\,s$^{-1}$ \citep{caswell82}. However, we do not detect any maser emission near these velocities.

Two of the  bar sources (354.701+0.299 and 354.724+0.300) are spatially and kinematically associated with the high velocity CO emission feature identified by \citet{bania77} and labelled as `Clump 1'. The compression and perturbation of Clump 1 would suggest star formation \citep[][and references therein]{fux99} and the presence of 6.7-GHz methanol maser emission confirms it. We do not detect any masers coincident with the second of Bania's CO structures, `Clump 2'. The high range of velocities ($>$100\,km\,s$^{-1}$) and small spatial extent of this clump, could be indicative of conditions too turbulent for the coherence necessary for maser emission.

\subsubsection{Orientation of the Bar}
The most densely populated portion of the far 3-kpc arm is between longitudes -10$^{\circ}$ and -15$^{\circ}$. This may correspond to enhanced high-mass star formation at the end of the Galactic bar, where the bar meets the 3-kpc arms \citep{caswell10mmb1}. This interpretation depends on the inclination of the Galactic bar to the Sun-Galactic centre line-of-sight as well as its length. Both of these are still a matter of debate, with the orientation varying between 15$^{\circ}$ and 45$^{\circ}$ \citep[e.g.][]{peters75, binney91, fux99, englmaier99, rodriguez08}, and the half-length between 2 and 5 kpc. However, the bar descriptions fall into two main categories: a short, bulge-like bar with an orientation of 20$^{\circ}$-25$^{\circ}$  \citep[e.g.][]{bissantz02, babusiaux05} and/or a long, thin bar with an orientation of $\sim$45$^{\circ}$  \citep[e.g.][]{hammersley00, benjamin05}. 

A bar orientation within the range 20$^{\circ}$-25$^{\circ}$ (and a half-length of 2.5 kpc) would imply the near-side bar end lying between 8$^{\circ}$ and 10$^{\circ}$ longitude and the far-side bar end lying between -4.5$^{\circ}$ and -6$^{\circ}$ longitude.  We find no corresponding high density regions: there are only three masers associated with the near-side and one with the far-side.
A bar orientation of approximately 45$^{\circ}$ (and a half-length of 3.5 kpc) would imply a near-side bar end at 22$^{\circ}$ longitude and a far-side bar end at -12$^{\circ}$ longitude. 
The extent of velocities for the near-side would depend on the nature of the 3-kpc arm structure beyond 15$^{\circ}$, but we do find a high density of masers at longitudes 18$^{\circ}$ to 20$^{\circ}$ with velocities between 10\,km\,s$^{-1}$ and 60\,km\,s$^{-1}$ (20 masers). The far-side is approximately coincident with the high density noted by \citet{caswell10mmb1}. 

%%%%%%%%%%%%%%%%%%%%%%%%%%%%%%%%%%%%%%%%%%%%%%%%%%%%%%%%%%%%

\begin{table} \centering \caption{\small Sources associated with the 3-kpc arms as identified by \citet{green09b} and \citet{caswell10mmb1}. For source details see: $^{1}$this paper; $^{2}$\citet{caswell10mmb1}. } 
\begin{tabular}{l c c r l} \\ \hline
Name (l, b) & V$_{pk}$ & V$_{mid}$  & S$_{pk}$ & Arm \\ 
 &  (km\,s$^{-1}$) & (km\,s$^{-1}$) & (Jy) &  \\ \hline  
000.212$-$0.001$^{2}$  & 49.5 & 45.8 & 2.40 & Far \\
002.143+0.009$^{2}$  & 62.7 & 59.5 & 6.70 & Far \\
003.442$-$0.348$^{2}$ & -35.2 & -35.0 & 0.66 & Near \\
005.618$-$0.082$^{2}$ & -27.1 & -23.3 & 3.42 & Near \\
006.189$-$0.358$^{1}$ & -30.2 & -32.3 & 221.60 & Near \\
007.166+0.131$^{1}$ & 85.7 & 83.1 & 2.47 & Far \\
008.832$-$0.028$^{1}$  & -3.8 & -1.6 & 126.80 & Near \\
009.619+0.193$^{1}$ & 7.0 & 5.5 & 5.5 &  Near \\
009.621+0.196$^{1}$ & 1.3 & 2.1 & 5196.00 & Near \\
010.629$-$0.333$^{1}$ & -0.4 & -6.4 & 4.20 & Near \\
010.724$-$0.334$^{1}$ & -2.1 & -2.1 & 1.51 & Near \\
012.025$-$0.025$^{1}$ & 108.3 & 109.2 & 103.10 & Far \\
013.696$-$0.156$^{1}$ & 99.4 & 102.0 & 1.86 & Far \\
014.101+0.087$^{1}$ & 15.4 & 10.5 & 86.55 & Near \\
014.521+0.155$^{1}$ & 4.1 & 3.7 & 1.31 & Near \\
345.198$-$0.030$^{2}$ & -0.5 & -1.5 & 2.23 & Far \\
345.441+0.205$^{2}$ & 0.9 & -5.5 & 2.32 & Far \\
345.505+0.348$^{2}$& -17.8 & -16.8 & 307.00 & Far \\
345.576$-$0.225$^{2}$& -126.8 & -124.6 & 0.65 & Near \\
345.807$-$0.044$^{2}$ & -2.0 & -1.8 & 1.21 & Far \\
345.824+0.044$^{2}$& -10.3 & -10.5 & 3.92 & Far \\
346.036+0.048$^{2}$& -6.4 & -9.2 & 10.42 & Far \\
346.481+0.132$^{2}$& -5.5 & -8.3 & 1.48 & Far \\
346.517+0.117$^{2}$ & -1.7 & -1.0 & 4.00 & Far \\
346.522+0.085$^{2}$& 5.7 & 5.4 & 1.47 & Far \\
347.583+0.213$^{2}$& -102.3 & -99.9 & 3.18 & Near \\
347.628+0.149$^{2}$& -96.5 & -96.9 & 18.98 & Near \\
348.027+0.106$^{2}$& -121.2 & -118.6 & 4.54 & Near \\
348.654+0.244$^{2}$& 16.9 & 17.0 & 0.99 & Far \\
348.723$-$0.078$^{2}$ & 11.5 & 10.5 & 2.25 & Far \\
348.892$-$0.180$^{2}$ & 1.5 & 1.5 & 2.44 & Far \\
349.067$-$0.017$^{2}$ & 11.6 & 11.0 & 2.43 & Far \\
349.151+0.021$^{2}$& 14.6 & 19.6 & 3.33 & Far \\
349.884+0.231$^{2}$ & 16.2 & 15.5 & 6.42 & Far \\
350.116+0.220$^{2}$ & 4.2 & 4.0 & 2.31 & Far \\
350.776+0.138$^{2}$ & 38.7 & 36.8 & 0.91 & Far \\
351.581$-$0.353$^{2}$ & -94.2 & -94.5 & 47.46 & Near \\
352.584$-$0.185$^{2}$ & -85.7 & -86.2 & 3.73 & Near \\
352.604$-$0.225$^{2}$ & -81.7 & -83.0 & 1.80 & Near \\
353.363$-$0.166$^{2}$ & -79.0 & -79.2 & 2.94 & Near \\
354.496+0.083$^{2}$ & 27.0 & 22.5 & 7.78 & Far \\
356.662$-$0.263$^{2}$ & -53.8 & -50.5 & 10.10 & Near \\
358.809$-$0.085$^{2}$ & -56.2 & -55.4 & 11.99 & Near \\
359.436$-$0.104$^{2}$ & -47.8 & -49.0 & 59.70 & Near \\
359.436$-$0.102$^{2}$ & -53.4 & -56.0 & 1.50 & Near \\
\hline
\end{tabular} 
\label{3-kpclist}
\end{table}

\subsection{Data Availability}

The data from the MMB survey is being made publicly available as each section
of the survey is published. The data are accessible from
http://www.jb.man.ac.uk/mmb or http://www.astromasers.org. 

\section{SUMMARY} 
We present the 119 sources detected by the Methanol Multibeam survey between longitudes 6$^{\circ}$ and 20$^{\circ}$, including 42 new detections. Consistent with expectations we find a narrow latitude distribution. We list the 45 sources within $\pm$15$^{\circ}$ of the Galactic centre associated with the 3-kpc arm features. We associate sources with the Galactic bar and relate the higher density regions of the 3-kpc arms to the interaction of a Galactic bar at an orientation of $\sim$45$^{\circ}$.

\section*{Acknowledgments} AA acknowledges the support of
a Science and Technology Facilities Council (STFC)  studentship. LQ
acknowledges the support of the EU Framework 6 Marie Curie Early Stage
Training programme under contract number MEST-CT-2005-19669 ``ESTRELA''.  
The Parkes Observatory and the Australia Telescope Compact
Array are part of the Australia Telescope which is funded by the
Commonwealth of Australia for operation as a National Facility managed by
CSIRO.

\bibliographystyle{mn2e} \bibliography{UberRef}

\begin{thebibliography}{}

\bibitem[\protect\citeauthoryear{{Anderson} \& {Bania}}{{Anderson} \&
  {Bania}}{2009}]{anderson09a}
{Anderson} L.~D.,  {Bania} T.~M.,  2009, ApJ, 690, 706

\bibitem[\protect\citeauthoryear{{Babusiaux} \& {Gilmore}}{{Babusiaux} \&
  {Gilmore}}{2005}]{babusiaux05}
{Babusiaux} C.,  {Gilmore} G.,  2005, MNRAS, 358, 1309

\bibitem[\protect\citeauthoryear{{Bania}}{{Bania}}{1977}]{bania77}
{Bania} T.~M.,  1977, ApJ, 216, 381

\bibitem[\protect\citeauthoryear{{Bania}}{{Bania}}{1980}]{bania80}
{Bania} T.~M.,  1980, ApJ, 242, 95

\bibitem[\protect\citeauthoryear{{Benjamin}, {Churchwell}, {Babler} \& {et
  al.}}{{Benjamin} et~al.}{2005}]{benjamin05}
{Benjamin} R.~A.,  {Churchwell} E.,  {Babler} B.~L.,    {et al.} 2005, ApJ
  Letters, 630, 149

\bibitem[\protect\citeauthoryear{{Beuther}, {Walsh}, {Schilke}, {Sridharan},
  {Menten} \& {Wyrowski}}{{Beuther} et~al.}{2002}]{beuther02}
{Beuther} H.,  {Walsh} A.,  {Schilke} P.,  {Sridharan} T.~K.,  {Menten} K.~M.,
    {Wyrowski} F.,  2002, A\&A, 390, 289

\bibitem[\protect\citeauthoryear{{Binney}, {Gerhard}, {Stark}, {Bally} \&
  {Uchida}}{{Binney} et~al.}{1991}]{binney91}
{Binney} J.,  {Gerhard} O.~E.,  {Stark} A.~A.,  {Bally} J.,    {Uchida} K.~I.,
  1991, MNRAS, 252, 210

\bibitem[\protect\citeauthoryear{{Bissantz} \& {Gerhard}}{{Bissantz} \&
  {Gerhard}}{2002}]{bissantz02}
{Bissantz} N.,  {Gerhard} O.,  2002, MNRAS, 330, 591

\bibitem[\protect\citeauthoryear{{Brand} \& {Blitz}}{{Brand} \&
  {Blitz}}{1993}]{brand93}
{Brand} J.,  {Blitz} L.,  1993, A\&A, 275, 67

\bibitem[\protect\citeauthoryear{{Caswell}}{{Caswell}}{1997}]{caswell97}
{Caswell} J.~L.,  1997, MNRAS, 289, 203

\bibitem[\protect\citeauthoryear{{Caswell}}{{Caswell}}{2009}]{caswell09a}
{Caswell} J.~L.,  2009, Publications of the Astronomical Society of Australia,
  26, 454

\bibitem[\protect\citeauthoryear{{Caswell}, {Fuller}, {Green} \& {et
  al.}}{{Caswell} et~al.}{2010}]{caswell10mmb1}
{Caswell} J.~L.,  {Fuller} G.~A.,  {Green} J.~A.,    {et al.} 2010, MNRAS, 404, 1029

\bibitem[\protect\citeauthoryear{{Caswell} \& {Haynes}}{{Caswell} \&
  {Haynes}}{1982}]{caswell82}
{Caswell} J.~L.,  {Haynes} R.~F.,  1982, ApJ Letters, 254, 31

\bibitem[\protect\citeauthoryear{{Caswell}, {Vaile}, {Ellingsen}, {Whiteoak} \&
  {Norris}}{{Caswell} et~al.}{1995a}]{caswell95a}
{Caswell} J.~L.,  {Vaile} R.~A.,  {Ellingsen} S.~P.,  {Whiteoak} J.~B.,
  {Norris} R.~P.,  1995a, MNRAS, 272, 96
  
\bibitem[\protect\citeauthoryear{{Caswell}, {Vaile} \& {Ellingsen}}{{Caswell}
  et~al.}{1995b}]{caswell95d}
{Caswell} J.~L.,  {Vaile} R.~A.,    {Ellingsen} S.~P.,  1995b, Publications of
  the Astronomical Society of Australia, 12, 37

\bibitem[\protect\citeauthoryear{{Cohen} \& {Davies}}{{Cohen} \&
  {Davies}}{1976}]{cohen76}
{Cohen} R.~J.,  {Davies} R.~D.,  1976, MNRAS, 175, 1

\bibitem[\protect\citeauthoryear{{Cyganowski}, {Brogan}, {Hunter} \&
  {Churchwell}}{{Cyganowski} et~al.}{2009}]{cyganowski09}
{Cyganowski} C.~J.,  {Brogan} C.~L.,  {Hunter} T.~R.,    {Churchwell} E.,
  2009, ApJ, 702, 1615

\bibitem[\protect\citeauthoryear{{Dame}, {Hartmann} \& {Thaddeus}}{{Dame}
  et~al.}{2001}]{dame01}
{Dame} T.~M.,  {Hartmann} D.,    {Thaddeus} P.,  2001, ApJ, 547, 792

\bibitem[\protect\citeauthoryear{{Dame} \& {Thaddeus}}{{Dame} \&
  {Thaddeus}}{2008}]{dame08}
{Dame} T.~M.,  {Thaddeus} P.,  2008, ApJ Letters, 683, 143

\bibitem[\protect\citeauthoryear{{Downes}, {Wilson}, {Bieging} \&
  {Wink}}{{Downes} et~al.}{1980}]{downes80b}
{Downes} D.,  {Wilson} T.~L.,  {Bieging} J.,    {Wink} J.,  1980, A\&As, 40,
  379

\bibitem[\protect\citeauthoryear{{Ellingsen}}{{Ellingsen}}{2007}]{ellingsen07}
{Ellingsen} S.~P.,  2007, MNRAS, 377, 571

\bibitem[\protect\citeauthoryear{{Englmaier} \& {Gerhard}}{{Englmaier} \&
  {Gerhard}}{1999}]{englmaier99}
{Englmaier} P.,  {Gerhard} O.,  1999, MNRAS, 304, 512

\bibitem[\protect\citeauthoryear{{Fish}, {Reid}, {Wilner} \&
  {Churchwell}}{{Fish} et~al.}{2003}]{fish03}
{Fish} V.~L.,  {Reid} M.~J.,  {Wilner} D.~J.,    {Churchwell} E.,  2003, ApJ,
  587, 701

\bibitem[\protect\citeauthoryear{{Fux}}{{Fux}}{1999}]{fux99}
{Fux} R.,  1999, A\&A, 345, 787

\bibitem[\protect\citeauthoryear{{Georgelin} \& {Georgelin}}{{Georgelin} \&
  {Georgelin}}{1976}]{georgelin76}
{Georgelin} Y.~M.,  {Georgelin} Y.~P.,  1976, A\&A, 49, 57

\bibitem[\protect\citeauthoryear{{Goedhart}, {Gaylard} \& {van der
  Walt}}{{Goedhart} et~al.}{2003}]{goedhart03}
{Goedhart} S.,  {Gaylard} M.~J.,    {van der Walt} D.~J.,  2003, MNRAS, 339,
  L33

\bibitem[\protect\citeauthoryear{{Goedhart}, {Gaylard} \& {van der
  Walt}}{{Goedhart} et~al.}{2004}]{goedhart04}
{Goedhart} S.,  {Gaylard} M.~J.,    {van der Walt} D.~J.,  2004, MNRAS, 355,
  553

\bibitem[\protect\citeauthoryear{{Goedhart}, {Langa}, {Gaylard} \& {van der
  Walt}}{{Goedhart} et~al.}{2009}]{goedhart09}
{Goedhart} S.,  {Langa} M.~C.,  {Gaylard} M.~J.,    {van der Walt} D.~J.,
  2009, MNRAS, 398, 995

\bibitem[\protect\citeauthoryear{{Green}, {Caswell}, {Fuller} \& {et
  al.,}}{{Green} et~al.}{2009a}]{green09a}
{Green} J.~A.,  {Caswell} J.~L.,  {Fuller} G.~A.,    {et al.,} 2009a, MNRAS,
  392, 783

\bibitem[\protect\citeauthoryear{{Green}, {McClure-Griffiths}, {Caswell},
  {Ellingsen}, {Fuller}, {Quinn} \& {Voronkov}}{{Green}
  et~al.}{2009b}]{green09b}
{Green} J.~A.,  {McClure-Griffiths} N.~M.,  {Caswell} J.~L.,  {Ellingsen}
  S.~P.,  {Fuller} G.~A.,  {Quinn} L.,    {Voronkov} M.~A.,  2009b, ApJ Letters,
  696, 156

\bibitem[\protect\citeauthoryear{{Hammersley}, {Garz{\'o}n}, {Mahoney},
  {L{\'o}pez-Corredoira} \& {Torres}}{{Hammersley} et~al.}{2000}]{hammersley00}
{Hammersley} P.~L.,  {Garz{\'o}n} F.,  {Mahoney} T.~J.,  {L{\'o}pez-Corredoira}
  M.,    {Torres} M.~A.~P.,  2000, MNRAS, 317, L45

\bibitem[\protect\citeauthoryear{{Hofner}, {Kurtz}, {Churchwell}, {Walmsley} \&
  {Cesaroni}}{{Hofner} et~al.}{1994}]{hofner94}
{Hofner} P.,  {Kurtz} S.,  {Churchwell} E.,  {Walmsley} C.~M.,    {Cesaroni}
  R.,  1994, ApJ Letters, 429, 85

\bibitem[\protect\citeauthoryear{{Kolpak}, {Jackson}, {Bania}, {Clemens} \&
  {Dickey}}{{Kolpak} et~al.}{2003}]{kolpak03}
{Kolpak} M.~A.,  {Jackson} J.~M.,  {Bania} T.~M.,  {Clemens} D.~P.,    {Dickey}
  J.~M.,  2003, ApJ, 582, 756
  
 \bibitem[\protect\citeauthoryear{{van der Kruit}}{{van der
  Kruit}}{1971}]{kruit71}
{van der Kruit} P.~C.,  1971, A\&A, 13, 405

\bibitem[Lockman(1980)]{lockman80} Lockman, F.J., 1980, ApJ, 241, 200

\bibitem[\protect\citeauthoryear{{MacLeod} \& {Gaylard}}{{MacLeod} \&
  {Gaylard}}{1992}]{macleod92b}
{MacLeod} G.~C.,  {Gaylard} M.~J.,  1992, MNRAS, 256, 519

\bibitem[\protect\citeauthoryear{{McClure-Griffiths}, {Dickey}, {Gaensler},
  {Green}, {Haverkorn} \& {Strasser}}{{McClure-Griffiths}
  et~al.}{2005}]{mcclure05}
{McClure-Griffiths} N.~M.,  {Dickey} J.~M.,  {Gaensler} B.~M.,  {Green} A.~J.,
  {Haverkorn} M.,    {Strasser} S.,  2005, ApJs, 158, 178

\bibitem[\protect\citeauthoryear{{McMillan} \& {Binney}}{{McMillan} \&
  {Binney}}{2010}]{mcmillan10}
{McMillan} P.~J.,  {Binney} J.~J.,  2010, MNRAS, 402, 934

\bibitem[\protect\citeauthoryear{{Menten}}{{Menten}}{1991}]{menten91}
{Menten} K.~M.,  1991, ApJ, 380, 75

\bibitem[\protect\citeauthoryear{{Minier}, {Ellingsen}, {Norris} \&
  {Booth}}{{Minier} et~al.}{2003}]{minier03}
{Minier} V.,  {Ellingsen} S.~P.,  {Norris} R.~P.,    {Booth} R.~S.,  2003,
  A\&A, 403, 1095

\bibitem[\protect\citeauthoryear{{Norris}, {Whiteoak}, {Caswell}, {Wieringa} \&
  {Gough}}{{Norris} et~al.}{1993}]{norris93}
{Norris} R.~P.,  {Whiteoak} J.~B.,  {Caswell} J.~L.,  {Wieringa} M.~H.,
  {Gough} R.~G.,  1993, ApJ, 412, 222
  
 \bibitem[\protect\citeauthoryear{{Oort}, {Kerr} \& {Westerhout}}{{Oort}
  et~al.}{1958}]{oort58}
{Oort} J.~H.,  {Kerr} F.~J.,    {Westerhout} G.,  1958, MNRAS, 118, 379

\bibitem[\protect\citeauthoryear{{Pandian}, {Menten} \& {Goldsmith}}{{Pandian}
  et~al.}{2009}]{pandian09}
{Pandian} J.~D.,  {Menten} K.~M.,    {Goldsmith} P.~F.,  2009, ArXiv e-prints

\bibitem[\protect\citeauthoryear{{Pestalozzi}, {Humphreys} \&
  {Booth}}{{Pestalozzi} et~al.}{2002}]{pestalozzi02b}
{Pestalozzi} M.,  {Humphreys} E.~M.~L.,    {Booth} R.~S.,  2002, A\&A, 384, 15

\bibitem[\protect\citeauthoryear{{Pestalozzi}, {Minier} \&
  {Booth}}{{Pestalozzi} et~al.}{2005}]{pestalozzi05}
{Pestalozzi} M.~R.,  {Minier} V.,    {Booth} R.~S.,  2005, A\&A, 432, 737

\bibitem[\protect\citeauthoryear{{Peters} III}{{Peters}}{1975}]{peters75}
{Peters} III W.~L.,  1975, ApJ, 195, 617

\bibitem[\protect\citeauthoryear{{Phillips}, {Norris}, {Ellingsen} \&
  {McCulloch}}{{Phillips} et~al.}{1998}]{phillips98}
{Phillips} C.~J.,  {Norris} R.~P.,  {Ellingsen} S.~P.,    {McCulloch} P.~M.,
  1998, MNRAS, 300, 1131

\bibitem[\protect\citeauthoryear{{Reid}, {Menten}, {Zheng}, {Brunthaler},
  {Moscadelli}, {Xu}, {Zhang}, {Sato}, {Honma}, {Hirota}, {Hachisuka}, {Choi},
  {Moellenbrock} \& {Bartkiewicz}}{{Reid} et~al.}{2009}]{reid09}
{Reid} M.~J.,  {Menten} K.~M.,  {Zheng} X.~W.,  {Brunthaler} A.,  {Moscadelli}
  L.,  {Xu} Y.,  {Zhang} B.,  {Sato} M.,  {Honma} M.,  {Hirota} T.,
  {Hachisuka} K.,  {Choi} Y.~K.,  {Moellenbrock} G.~A.,    {Bartkiewicz} A.,
  2009, ApJ, 700, 137

\bibitem[\protect\citeauthoryear{{Rodriguez-Fernandez} \&
  {Combes}}{{Rodriguez-Fernandez} \& {Combes}}{2008}]{rodriguez08}
{Rodriguez-Fernandez} N.~J.,  {Combes} F.,  2008, A\&A, 489, 115

\bibitem[\protect\citeauthoryear{{Roman-Duval}, {Jackson}, {Heyer}, {Johnson},
  {Rathborne}, {Shah} \& {Simon}}{{Roman-Duval} et~al.}{2009}]{roman09}
{Roman-Duval} J.,  {Jackson} J.~M.,  {Heyer} M.,  {Johnson} A.,  {Rathborne}
  J.,  {Shah} R.,    {Simon} R.,  2009, ApJ, 699, 1153

\bibitem[\protect\citeauthoryear{{Rygl}, {Brunthaler}, {Reid}, {Menten}, {van
  Langevelde} \& {Xu}}{{Rygl} et~al.}{2010}]{rygl10}
{Rygl} K.~L.~J.,  {Brunthaler} A.,  {Reid} M.~J.,  {Menten} K.~M.,  {van
  Langevelde} H.~J.,    {Xu} Y.,  2010, A\&A, 511, A2+

\bibitem[\protect\citeauthoryear{{Sanna}, {Reid}, {Moscadelli}, {Dame},
  {Menten}, {Brunthaler}, {Zheng} \& {Xu}}{{Sanna} et~al.}{2009}]{sanna09}
{Sanna} A.,  {Reid} M.~J.,  {Moscadelli} L.,  {Dame} T.~M.,  {Menten} K.~M.,
  {Brunthaler} A.,  {Zheng} X.~W.,    {Xu} Y.,  2009, ApJ, 706, 464

\bibitem[\protect\citeauthoryear{{Schutte}, {van der Walt}, {Gaylard} \&
  {MacLeod}}{{Schutte} et~al.}{1993}]{schutte93}
{Schutte} A.~J.,  {van der Walt} D.~J.,  {Gaylard} M.~J.,    {MacLeod} G.~C.,
  1993, MNRAS, 261, 783

\bibitem[\protect\citeauthoryear{{Sevenster}, {Saha}, {Valls-Gabaud} \&
  {Fux}}{{Sevenster} et~al.}{1999}]{sevenster99}
{Sevenster} M.,  {Saha} P.,  {Valls-Gabaud} D.,    {Fux} R.,  1999, MNRAS, 307,
  584

\bibitem[\protect\citeauthoryear{{Sewilo}, {Watson}, {Araya}, {Churchwell},
  {Hofner} \& {Kurtz}}{{Sewilo} et~al.}{2004}]{sewilo04}
{Sewilo} M.,  {Watson} C.,  {Araya} E.,  {Churchwell} E.,  {Hofner} P.,
  {Kurtz} S.,  2004, ApJs, 154, 553

\bibitem[\protect\citeauthoryear{{Slysh}, {Val'tts}, {Kalenskii}, {Voronkov},
  {Palagi}, {Tofani} \& {Catarzi}}{{Slysh} et~al.}{1999}]{slysh99}
{Slysh} V.~I.,  {Val'tts} I.~E.,  {Kalenskii} S.~V.,  {Voronkov} M.~A.,
  {Palagi} F.,  {Tofani} G.,    {Catarzi} M.,  1999, A\&AS, 134, 115

\bibitem[\protect\citeauthoryear{{Solomon}, {Rivolo}, {Barrett} \&
  {Yahil}}{{Solomon} et~al.}{1987}]{solomon87}
{Solomon} P.~M.,  {Rivolo} A.~R.,  {Barrett} J.,    {Yahil} A.,  1987, ApJ,
  319, 730

\bibitem[\protect\citeauthoryear{{Szymczak}, {Bartkiewicz} \&
  {Richards}}{{Szymczak} et~al.}{2007}]{szymczak07}
{Szymczak} M.,  {Bartkiewicz} A.,    {Richards} A.~M.~S.,  2007, A\&A, 468, 617

\bibitem[\protect\citeauthoryear{{Szymczak}, {Hrynek} \& {Kus}}{{Szymczak}
  et~al.}{2000}]{szymczak00}
{Szymczak} M.,  {Hrynek} G.,    {Kus} A.~J.,  2000, A\&AS, 143, 269

\bibitem[\protect\citeauthoryear{{Taylor} \& {Cordes}}{{Taylor} \&
  {Cordes}}{1993}]{taylor93}
{Taylor} J.~H.,  {Cordes} J.~M.,  1993, ApJ, 411, 674

\bibitem[\protect\citeauthoryear{{van der Walt}, {Gaylard} \& {MacLeod}}{{van
  der Walt} et~al.}{1995}]{walt95}
{van der Walt} D.~J.,  {Gaylard} M.~J.,    {MacLeod} G.~C.,  1995, A\&AS, 110,
  81

\bibitem[\protect\citeauthoryear{{van der Walt}, {Goedhart} \& {Gaylard}}{{van
  der Walt} et~al.}{2009}]{walt09}
{van der Walt} D.~J.,  {Goedhart} S.,    {Gaylard} M.~J.,  2009, MNRAS, 398,
  961

\bibitem[\protect\citeauthoryear{{Vlemmings}, {Goedhart} \&
  {Gaylard}}{{Vlemmings} et~al.}{2009}]{vlemmings09}
{Vlemmings} W.~H.~T.,  {Goedhart} S.,    {Gaylard} M.~J.,  2009, A\&A, 500, L9

\bibitem[\protect\citeauthoryear{{Voronkov}, {Caswell}, {Ellingsen} \&
  {Sobolev}}{{Voronkov} et~al.}{2010}]{voronkov10}
{Voronkov} M.~A.,  {Caswell} J.~L.,  {Ellingsen} S.~P.,    {Sobolev} A.~M.,
  2010, 405, 2471

\bibitem[\protect\citeauthoryear{{Walsh}, {Burton}, {Hyland} \&
  {Robinson}}{{Walsh} et~al.}{1998}]{walsh98}
{Walsh} A.~J.,  {Burton} M.~G.,  {Hyland} A.~R.,    {Robinson} G.,  1998,
  MNRAS, 301, 640

\bibitem[\protect\citeauthoryear{{Walsh}, {Hyland}, {Robinson} \&
  {Burton}}{{Walsh} et~al.}{1997}]{walsh97}
{Walsh} A.~J.,  {Hyland} A.~R.,  {Robinson} G.,    {Burton} M.~G.,  1997,
  MNRAS, 291, 261

\bibitem[\protect\citeauthoryear{{Wilson}, {Walmsley}, {Jewell} \&
  {Snyder}}{{Wilson} et~al.}{1984}]{wilson84}
{Wilson} T.~L.,  {Walmsley} C.~M.,  {Jewell} P.~R.,    {Snyder} L.~E.,  1984,
  A\&A, 134, L7

\bibitem[\protect\citeauthoryear{{Wink}, {Altenhoff} \& {Mezger}}{{Wink}
  et~al.}{1982}]{wink82}
{Wink} J.~E.,  {Altenhoff} W.~J.,    {Mezger} P.~G.,  1982, A\&A, 108, 227

\bibitem[\protect\citeauthoryear{{van Woerden}, {Rougoor} \& {Oort}}{{van
  Woerden} et~al.}{1957}]{vanwoerden57}
{van Woerden} H.,  {Rougoor} G.~W.,    {Oort} J.~H.,  1957, Academie des
  Sciences Paris Comptes Rendus, 244, 1691

\bibitem[\protect\citeauthoryear{{Xu}, {Li}, {Hachisuka}, {Pandian}, {Menten}
  \& {Henkel}}{{Xu} et~al.}{2008}]{xu08}
{Xu} Y.,  {Li} J.~J.,  {Hachisuka} K.,  {Pandian} J.~D.,  {Menten} K.~M.,
  {Henkel} C.,  2008, A\&A, 485, 729

\bibitem[\protect\citeauthoryear{{Xu}, {Voronkov}, {Pandian}, {Li}, {Sobolev},
  {Brunthaler}, {Ritter} \& {Menten}}{{Xu} et~al.}{2009}]{xu09b}
{Xu} Y.,  {Voronkov} M.~A.,  {Pandian} J.~D.,  {Li} J.~J.,  {Sobolev} A.~M.,
  {Brunthaler} A.,  {Ritter} B.,    {Menten} K.~M.,  2009, A\&A, 507, 1117

\end{thebibliography}

\label{lastpage}

\end{document}